\documentclass[structabstract]{aa}

\usepackage{amsfonts}
\usepackage{times}
\usepackage{xspace,amssymb,amsmath,tabularx}

\usepackage{color}
\usepackage[bookmarksopen=false,bookmarksnumbered=true,breaklinks=true,
            colorlinks=true,linkcolor=blue,citecolor=blue]{hyperref}

\usepackage{graphicx}

\DeclareGraphicsExtensions{.eps}
\usepackage{multirow}
\usepackage{array}

\begin{document}
\title{Coronagraphic phase diversity: performance study and laboratory demonstration}

\author{B. Paul\inst{1}\inst{2}\inst{3} \and J.-F. Sauvage\inst{1}\inst{3} \and L. M.
  Mugnier\inst{1}\inst{3}
}

\institute{Onera - The French Aerospace Lab, F-92322 Chatillon France
\and Aix Marseille Universit\'e, CNRS, LAM (Laboratoire d'Astrophysique de Marseille) UMR 7326, 13388, Marseille, France
\and Groupement d'int\'er\^et scientifique PHASE (Partenariat Haute
r\'esolution Angulaire Sol et Espace) between Onera, Observatoire de Paris, 
CNRS, Universit\'e Diderot, Laboratoire d'Astrophysique de Marseille and
Institut de Plan\'etologie et d'Astrophysique de Grenoble}

\date{Preprint online version: March 1, 2013}

\abstract {The final performance of current and future instruments dedicated
  to exoplanet detection and characterization (such as SPHERE on the European
  Very Large Telescope, GPI on Gemini North, or future instruments on
  Extremely Large Telescopes) is limited by uncorrected quasi-static
  aberrations. These aberrations create long-lived speckles in the scientific
  image plane, which can easily be mistaken for planets.} 
{Common adaptive optics systems require dedicated components to perform
  wave-front analysis. The ultimate wave-front measurement performance is thus
  limited by the unavoidable differential aberrations between the wavefront
  sensor and the scientific camera. To reach the level of detectivity required
  by high-contrast imaging, these differential aberrations must be estimated
  and compensated for. In this paper, we characterize and experimentally
  validate a wave-front sensing method that relies on focal-plane data.}
{Our method, called COFFEE (for COronagraphic Focal-plane wave-Front
  Estimation for Exoplanet detection), is based on a Bayesian approach, and it
  consists in an extension of phase diversity to high-contrast imaging. It
  estimates the differential aberrations using only two focal-plane
  coronagraphic images recorded from the scientific camera itself.}
{We first present a thorough characterization of COFFEE's performance by means
  of numerical simulations. This characterization is then compared with an
  experimental validation of COFFEE using an in-house adaptive optics bench
  and an apodized Roddier \& Roddier phase mask coronagraph. An excellent
  match between experimental results and the theoretical study is found.
  Lastly, we present a preliminary validation of COFFEE's ability to
  compensate for the aberrations upstream of a coronagraph.}
{}

\keywords{instrumentation: adaptive optics, instrumentation: high angular
  resolution, techniques: image processing, methods: numerical, methods:
  laboratory, telescopes}

\maketitle

\section{Introduction}

Exoplanet imaging is one of the main challenges in today's astronomy. A direct
observation of these planets can provide information on both the chemical
composition of their atmospheres and their temperatures. Such observations
have recently been made possible \citep{exop_kalas, exop_marois,
  Lagrange-a-09}, but only thanks to their high mass or their wide apparent
distance from their host
star.\\
Being able to image an object as faint as an extra-solar planet very close to
its parent star requires the use of extreme AO (XAO) systems coupled to a
high-contrast imaging technique, such as coronagraphy. Instruments dedicated
to exoplanet imaging using these two techniques (SPHERE on the VLT,
\citep{Beuzit-p-07}, GPI on Gemini North, \citep{gpi}) are currently being
integrated. The performance of such systems is limited by residual speckles on
the detector. These speckles, which originate in quasi-static non common path
aberrations (NCPA), strongly decrease the extinction provided by the
coronagraph and can be difficult to distinguish from an exoplanet. To achieve
the ultimate system performance, these aberrations must be measured and
compensated for. The current-generation instruments, SPHERE and GPI,
respectively rely on phase diversity \citep{Gonsalves} and an interferometry
approach
\citep{gpi_wallace_2010} to compensate for these NCPA.\\
Several techniques dedicated to high-contrast imaging system optimization have
been proposed for future systems. Some of them rely on a dedicated wave-front
sensing hardware \citep{clowfs}, others use scientific focal plane data
assuming small aberrations. Speckle nulling iterative techniques
\citep{speckle_nulling, efc} estimate the electric field in the detector plane
using at least three images. The technique proposed by \cite{scc} relies on a
modification of the imaging system, but requires only one image. These
techniques aim at minimizing the energy in a chosen area (``Dark Hole''),
leading to a contrast optimization on the detector \citep{spie_2007_jtrauger,
  spie_2012_pbaudoz} in a closed
loop process.\\
We have recently proposed a focal-plane wave-front sensor, COFFEE
\citep{Sauvage-a-12}, which is an extension of conventional phase diversity
\citep{Mugnier-l-06a} to a coronagraphic system. Since COFFEE uses
focal-plane images, it is possible to characterize the whole bench without any
differential aberration. This method requires only two focal-plane images to
estimate the aberrations upstream of the coronagraph without any modification
of the coronagraphic imaging system or assuming small aberrations. COFFEE's
principle and its application to the apodized Roddier \& Roddier phase mask
(ARPM) are described in Section \ref{coffee_principe}. In Section
\ref{coffee_param}, we evaluate the quality of NCPA estimation by realistic
simulations. In Section \ref{coffee_boa}, we present the experimental results
from the laboratory demonstration of COFFEE on an in-house adaptive optics
bench (BOA) with an ARPM. Section \ref{ccl} concludes the paper.

\section{COFFEE: principle}
\label{coffee_principe}

\subsection{Extension of phase diversity to coronagraphic images}
\label{coffee_principe_cpd}

\begin{figure}
\centering
\scalebox{0.25}{\input{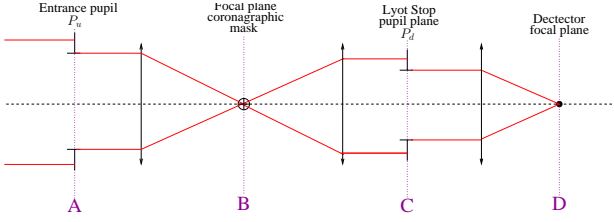}}
\caption{Coronagraphic imaging instrument: principle}
\label{syst_coro}
\end{figure}

Figure \ref{syst_coro} describes the coronagraphic imaging scheme considered
in this paper. We consider four successive planes denoted A (circular entrance
pupil of diameter $D_u$), B (coronagraphic focal plane), C (Lyot Stop), and D
(detector plane). The optical aberrations are considered as static and
introduced in the pupil planes A and C. The coronagraphic device is composed
of a focal plane mask located in plane B and a Lyot Stop situated in plane C.
No particular assumption is made on the pupil shape or intensity. Thus, the
description of COFFEE is compatible with several coronagraphic devices. COFFEE
uses two images, $\boldsymbol{i}_c^\text{f}$ and $\boldsymbol{i}_c^\text{d}$,
recorded on the detector (plane D in Figure \ref{syst_coro}) that, as in phase
diversity, differ from a known aberration, $\boldsymbol{\phi}_{div}$, to
estimate aberrations both upstream ($\boldsymbol{\phi}_u$) and downstream
($\boldsymbol{\phi}_d$) of the coronagraph.
 
Considering the calibration of the instrument with an unresolved object, we
use the following imaging model:
\begin{equation}\label{eq_im_model}
\begin{aligned}
\boldsymbol{i}_c^{\text{foc}}&=\alpha.\boldsymbol{h}_{\text{det}}\star\boldsymbol{h}_c(\boldsymbol{\phi}_u,\boldsymbol{\phi}_d)+\boldsymbol{n}^{\text{foc}}+\beta\\
\boldsymbol{i}_c^{\text{div}}&=\alpha.\boldsymbol{h}_{\text{det}}\star \boldsymbol{h}_c(\boldsymbol{\phi}_u+\boldsymbol{\phi}_{div},\boldsymbol{\phi}_d)+\boldsymbol{n}^{\text{div}}+\beta
\end{aligned}
\end{equation}
where $\alpha$ is the incoming flux, $\boldsymbol{h}_c$ the coronagraphic
``point spread function'' (PSF) of the instrument (i.e. the response of a
coronagraphic imaging system to a point source), $\boldsymbol{h}_{\text{det}}$
the known detector PSF, $\boldsymbol{n}^{\text{foc}}$ and
$\boldsymbol{n}^{\text{div}}$ are the measurement noises, $\beta$ is a uniform
background (offset), and $\star$ denotes the discrete convolution operation.
Such an imaging model can be used for any coronagraphic PSF expression
$\boldsymbol{h}_c$. The measurement noises $\boldsymbol{n}^{\text{foc}}$ and
$\boldsymbol{n}^{\text{div}}$ comprise both photon and detector noises.
Because calibration is assumed to be performed with high flux levels, we adopt
a non-stationary white Gaussian model, which is a good approximation of a mix
of photon and detector noises. Its variance is the sum of the photon and
detector noise variances:
$\boldsymbol{\sigma}^2_n[t]=\boldsymbol{\sigma}^2_{\text{ph}}[t] +
\sigma^2_{\text{det}}$ \citep{Mugnier-a-04}, with $t$ the pixel position in
the detector plane. The former can be estimated as the image itself
thresholded to positive values, and the latter can be calibrated prior to the
observations.

We adopt a maximum \textit{a posteriori} (MAP) approach and estimate the
aberrations $\phi_u$ and $\phi_d$, the flux $\alpha$, and the background
$\beta$ that minimize the neg-log-likelihood of the data, potentially
penalized by regularization terms on $\boldsymbol{\phi}_u$ and
$\boldsymbol{\phi}_d$ designed to enforce smoothness of the sought phases:
\begin{equation}
(\hat{\alpha}, \hat{\beta}, \hat{\boldsymbol{\phi}_u}, \hat{\boldsymbol{\phi}_d}) = 
\underset{\alpha, \beta, \boldsymbol{\phi}_u, \boldsymbol{\phi}_d}{\arg \min} J(\alpha, \beta, \boldsymbol{\phi}_u, \boldsymbol{\phi}_d )
\end{equation}
where
\begin{equation}\label{eq-pb-inverse}
\begin{aligned}
J(\alpha, \beta, \boldsymbol{\phi}_u, \boldsymbol{\phi}_d)&=\frac{1}{2}\left\|\frac{\boldsymbol{i}_c^{\text{foc}} 
- (\alpha.\boldsymbol{h}_{\text{d}}\star\boldsymbol{h}_c(\boldsymbol{\phi}_u,\boldsymbol{\phi}_d)+\beta)}{\boldsymbol{\sigma}_n^{\text{foc}}} \right\|^2 \\
&+\frac{1}{2}\left\|\frac{\boldsymbol{i}_c^{\text{div}} 
- (\alpha.\boldsymbol{h}_{\text{d}}\star\boldsymbol{h}_c(\boldsymbol{\phi}_u+\boldsymbol{\phi}_{div},\boldsymbol{\phi}_d)+\beta)}{\boldsymbol{\sigma}_n^{\text{div}}}\right\|^2\\
&+\mathcal{R}(\boldsymbol{\phi}_u) + \mathcal{R}(\boldsymbol{\phi}_d)
\end{aligned}
\end{equation}
where $\left\|{\boldsymbol{x}}\right\|^2$ denotes the sum of squared pixel
values of map $\boldsymbol{x}$, $\boldsymbol{\sigma}_n^{\text{foc}}$, and
$\boldsymbol{\sigma}_{n}^{\text{div}}$ are the noise standard deviation maps
of each image,
and $\mathcal{R}$ is a regularization metric for the phase. \\
Any aberration $\boldsymbol{\phi}$ is expanded on a basis
$\{\boldsymbol{Z}_k\}$ that is typically either Zernike polynomials or the
pixel indicator functions in the corresponding pupil plane:
$\boldsymbol{\phi}=\sum_k a_k\boldsymbol{Z}_k$ where the summation is, in
practice, limited to the number of coefficients considered sufficient to
correctly describe the aberrations. In this paper, the phase is expanded on a
truncated Zernike basis. The impact of using a regularization metric with such
a basis is studied later in this paper. In the MAP framework, the
regularization metrics $\mathcal{R}(\boldsymbol{\phi}_u)$ and
$\mathcal{R}(\boldsymbol{\phi}_d)$ are deduced from the assumed \textit{a
  priori} statistics of $\boldsymbol{\phi}_u$ and $\boldsymbol{\phi}_d$.
Assuming these aberrations are zero-mean, Gaussian, and neglecting \textit{a
  priori} correlations between Zernike modes, we obtain, for an estimation
performed on $N$ Zernike modes:
\begin{equation}\label{eq_regul}
  \mathcal{R}(\boldsymbol{\phi}_x)=\frac{1}{2}\boldsymbol{a}_x^tR_{a_x}^{-1}\boldsymbol{a}_x = \frac{1}{2}\sum_{k=1}^N\frac{a_{x_k}^2}{\sigma_{x_k}^2}\text{,}
\end{equation} 
where $\sigma_{x_k}^2$ is the assumed phase variance per Zernike mode,
$R_{a_k}$ the covariance matrix, and $\boldsymbol{a}_x$ a $N$ element vector
containing the estimated Zernike coefficients $a_{x_k}$. Here $x$ is either $u$ (upstream) or $d$ (downstream).\\
The minimization of metric $J(\alpha, \beta, \boldsymbol{\phi}_u,
\boldsymbol{\phi}_d)$ of Eq.\eqref{eq-pb-inverse} is performed by means of a
limited memory variable metric (BFGS) method \citep{numerical_recipes,
  Thiebaut-p-02}, which is a fast quasi-Newton type minimization method. It
uses both gradients $\frac{\partial J}{\partial\boldsymbol{\phi}_u}$ and
$\frac{\partial J}{\partial\boldsymbol{\phi}_d}$. Flux $\alpha$ and offset
$\beta$ are analytically obtained using gradients $\frac{\partial J}{\partial
  \alpha}$ and $\frac{\partial J}{\partial \beta}$ (implementation details,
including gradient expressions, can be found in Appendix \ref{impl_details}).

\cite{Sauvage-a-12} established that a suitable diversity phase
$\boldsymbol{\phi}_{div}$ for COFFEE was a mix of defocus and
astigmatism:
$\boldsymbol{\phi}_{div}=a_4^{div}\boldsymbol{Z}_4+a_5^{div}\boldsymbol{Z}_5$
with $a_4^{div}=a_5^{div}=80\text{ nm RMS}$, introduced upstream of the
coronagraph. We therefore use this diversity phase in the following.

\subsection{Coronagraphic imaging model}
\label{coffee_principe_im_model}

The imaging model used by COFFEE in the criterion minimization (equation
\eqref{eq-pb-inverse}) requires a coronagraphic PSF expression. In this paper,
we use the analytical coronagraphic imaging model developed by
\cite{Sauvage-a-10}, whose formalism is developed in this section, where
$\boldsymbol{r}$ is the pupil plane position vector, $r$ its modulus, and
$\boldsymbol{\gamma}$ the focal plane position vector. The entrance pupil
function $\boldsymbol{P}_u(\boldsymbol{r})$ is such that:
\begin{equation}\label{pup_in_trans}
\boldsymbol{P}_u(\boldsymbol{r})=\boldsymbol{\Pi}\left(\frac{2r}{D_u}\right)\boldsymbol{\Phi}(\boldsymbol{r})
\end{equation}
with $\boldsymbol{\Pi}\left(\frac{2r}{D_u}\right) = 1$ for $r\leq
\frac{D_u}{2}$, pupil entrance diameter, $0$ otherwise, and
$\boldsymbol{\Phi}$ is an apodization function. In this paper, we consider
that the impact of amplitude aberrations is negligible, which is a reasonable
assumption for a ground-based, high-contrast imaging system such as SPHERE.
Considering only static aberrations (no residual turbulent aberrations), the
electric field $\boldsymbol{\Psi}_A$ in the entrance pupil plane can be
written as
\begin{equation}
\boldsymbol{\Psi}_A(\boldsymbol{r})=\boldsymbol{P}_u(\boldsymbol{r})e^{j\boldsymbol{\phi}_u(\boldsymbol{r})}\text{,}
\end{equation}
The field amplitude $\boldsymbol{\Psi}_B(\boldsymbol{\gamma})$ in plane B can be
calculated, following \cite{Sauvage-a-10}, using the analytical
coronagraphic imaging model (which is called ``perfect coronagraph
model'' hereafter):
\begin{equation}\label{eq_coro_B}
  \boldsymbol{\Psi}_B(\boldsymbol{\gamma})=\text{FT}^{-1}(\boldsymbol{\Psi}_A(\boldsymbol{r}))-\eta_0\text{FT}^{-1}(\boldsymbol{P}_u(\boldsymbol{r}))\text{,}
\end{equation}
where $\eta_0$ is the scalar that minimizes the outcoming energy from focal plane B,
whose analytical value is given by
\begin{equation}\label{eta0}
\eta_0=\frac{1}{\mathcal{N}}\iint_{S}\boldsymbol{\Psi}_A^*(\boldsymbol{r})\boldsymbol{P}_u(\boldsymbol{r})d\boldsymbol{r}\text{,}
\end{equation}
where
\begin{equation}
\mathcal{N}=\iint_{S}\boldsymbol{P}_u^*(\boldsymbol{r})\boldsymbol{P}_u(\boldsymbol{r})d\boldsymbol{r}.
\end{equation}

It is worthy mentioning that $\eta_0$ is the exact definition of the
instantaneous Strehl ratio given by \cite{Born_Wolf}. One can notice that
$\eta_0=1$ when there is no aberration upstream of the coronagraph
($\boldsymbol{\phi}_u(\boldsymbol{r})=0$), so that $\boldsymbol{\Psi}_B=0$ in
such a case. No aberration in the entrance pupil leads to no outcoming energy
from plane B, and thus to a perfect extinction in the
detector plane D.\\
Propagating the wave from plane B (Eq. \eqref{eq_coro_B}) to plane D, we can
write the electric field $\boldsymbol{\Psi}_D(\boldsymbol{\gamma})$ in the
detector plane:
\begin{equation}
\begin{aligned}
\boldsymbol{\Psi}_D(\boldsymbol{\gamma})=&\text{FT}^{-1}\left\{\boldsymbol{P}_d(\boldsymbol{r})e^{j(\boldsymbol{\phi}_u(\boldsymbol{r})+\boldsymbol{\phi}_d(\boldsymbol{r}))}\right\}\\
&-\eta_0\text{FT}^{-1}\left\{\boldsymbol{P}_d(\boldsymbol{r})e^{j\boldsymbol{\phi}_d(\boldsymbol{r})}\right\}\text{,}
\end{aligned}
\end{equation}
where $\boldsymbol{P}_d(\boldsymbol{r})$ is the Lyot stop pupil function:
$\boldsymbol{P}_d(\boldsymbol{r})=\boldsymbol{\Pi}\left(\frac{2r}{D_d}\right)\boldsymbol{P}_u(\boldsymbol{r})$, with $D_d$ the Lyot stop pupil
diameter ($D_d\leq D_u$). 
For the sake of simplicity, we omit the spatial variables $\boldsymbol{r}$
and $\boldsymbol{\gamma}$ in the following. The coronagraphic PSF of
the instrument, denoted by $\boldsymbol{h}_c$, is the square modulus of $\boldsymbol{\Psi}_D$:
\begin{equation}\label{eq_coro_prf}
\begin{aligned}
\boldsymbol{h}_c(\boldsymbol{\phi}_u,\boldsymbol{\phi}_d)=&\big|\text{FT}^{-1}(\boldsymbol{P}_de^{j(\boldsymbol{\phi}_u+\boldsymbol{\phi}_d)})\\
&-\eta_0\text{FT}^{-1}(\boldsymbol{P}_de^{j\boldsymbol{\phi}_d})\big|^2\text{.}
\end{aligned}
\end{equation}

In this paper, this expression of the coronagraphic PSF is the one used by
COFFEE for estimating $\boldsymbol{\phi}_u$ and $\boldsymbol{\phi}_d$; i.e.,
Eq.~\eqref{eq_coro_prf} is inserted into the imaging model
(Eq.~\eqref{eq_im_model}) used in the criterion
minimization described in Eq.~\eqref{eq-pb-inverse}.\\
As described by \cite{Sauvage-a-10}, this model, which analytically describes
the impact of a coronagraph in an imaging system, considers that the
coronagraph removes the projection of the incoming electric field on an Airy
pattern, represented by the parameter $\eta_0$ (Eq.~\eqref{eta0}). Since it
does not assume small aberrations, it can be used for any wave-front error
upstream of the coronagraph. The quality of the fit of this analytical imaging
model with the ARPM coronagraph is discussed later in this paper (Section
\ref{coffee_param_model}).

\section{Performance assessment by numerical simulation}
\label{coffee_param}

The aim of this section is to quantify the impact of each error source on
COFFEE's aberration estimation. Such a study will show COFFEE's sensitivity to
the classical error sources that limit the phase retrieval in a real system
(and thus the final extinction of the coronagraph), which will be of high
interest in defining COFFEE's upgrades. Likewise, it will allow us to estimate
the accuracy level expected on our AO bench. In this section, we present the
evolution of this reconstruction error with respect to the incoming flux
(Section \ref{coffee_param_snr}), to the size of the source (Section
\ref{coffee_param_obj}), to an error made on the assumed diversity phase used
in the reconstruction (Section \ref{coffee_param_errdiv}), and to the number
of Zernike modes used in the reconstruction (Section
\ref{coffee_param_aliasing}). For each error source, coronagraphic images will
be computed using the imaging model presented in Eq. \eqref{eq_im_model},
using the perfect coronagraph model to calculate the coronagraphic PSF $h_c$
whose expression is given Eq. \eqref{eq_coro_prf}. COFFEE will then perform
the phase estimation using these two images. The compatibility of COFFEE with
realistic coronagraphic images will be studied as well (Section
\ref{coffee_param_model}) by computing coronagraphic images using a realistic
coronagraph model and then running COFFEE to estimate the aberrations both
upstream and
downstream of the coronagraph.\\
Table \ref{table_snr_obj_errdiv} gathers the parameters used for these
simulations.

\begin{tiny}
\begin{table}
\begin{tabular}{m{4cm} m{4cm}}
  \hline
  \multicolumn{2}{c}{\textbf{Simulation}}\\
  \hline
  image size & $93\times 93$ $\frac{\lambda}{D}$ ($128\times128$ pixels, oversampling factor: $1.38$)\\
  Light spectrum & monochromatic ($\lambda \!=\! 635$ nm)\\
  Aberration upstream of the coronagraph ($\boldsymbol{\phi}_u$) & WFE $=80$ nm RMS\\
  Aberration downstream of the coronagraph ($\boldsymbol{\phi}_d$) & WFE $=20$ nm RMS\\  
  Zernike basis used for $\boldsymbol{\phi}_u$ and $\boldsymbol{\phi}_d$ simulation & $36$ Zernike polynomials\\
  \hline
  \hline
  \multicolumn{2}{c}{\textbf{Phase estimation: COFFEE}}\\
  \hline
  Zernike basis used for $\boldsymbol{\phi}_u$ and $\boldsymbol{\phi}_d$ reconstruction & $36$ Zernike polynomials\\
  Regularization metric & none\\
  \hline
\end{tabular}
\caption{COFFEE: simulation parameters used for the performance assessments of
sections \ref{coffee_param_snr}, \ref{coffee_param_obj} and \ref{coffee_param_errdiv}}
\label{table_snr_obj_errdiv}
\end{table}
\end{tiny}

The chosen wave-front error (WFE) values upstream and downstream of the
coronagraph for these simulations are typical of the aberrations that will be
estimated on our AO bench in Section \ref{coffee_boa} (so that experimental
results can be compared to the following simulations). Since these simulations
are performed with a small number of Zernike modes ($36$), there is no need of
regularization metrics in such simulations.\\
To simulate realistic aberrations, we have considered that the variance per
Zernike mode $\sigma_k^2$ was decreasing with the radial order $n(k)$ of the
considered Zernike mode $k$ \citep{Noll-a-76}:
\begin{equation}\label{phi_spectrum}
\sigma_k^2 \propto \frac{1}{n(k)^2}\ \text{.}
\end{equation}
This corresponds to a decrease in the static aberration spatial spectrum as
$\frac{1}{|\nu|^2}$, where $\nu$ is the spatial frequency, which is a common
assumption for mirror fabrication errors. To evaluate COFFEE's performance, we
define the reconstruction error $\epsilon_x$ ($x$ stands for $u$ (upstream) or
$d$ (downstream)) as
\begin{equation}\label{eq_err_rec}
\epsilon=\sqrt{\sum_{k=2}^{N-1}|a_k-\hat{a}_k|^2}
\end{equation}
with $a_k$ the Zernike coefficients (starting with $k=2$ corresponding to
tilt) used for the simulation, $\hat{a}_k$ the reconstructed Zernike
coefficients, and $N$ the number of Zernike modes. In this section, every
reconstruction error value is an average value, computed from ten independent
simulated phases.

\subsection{Noise propagation}
\label{coffee_param_snr}
\begin{figure}
\centering
\begin{tabular}{c}
\includegraphics[width = 0.8\linewidth]{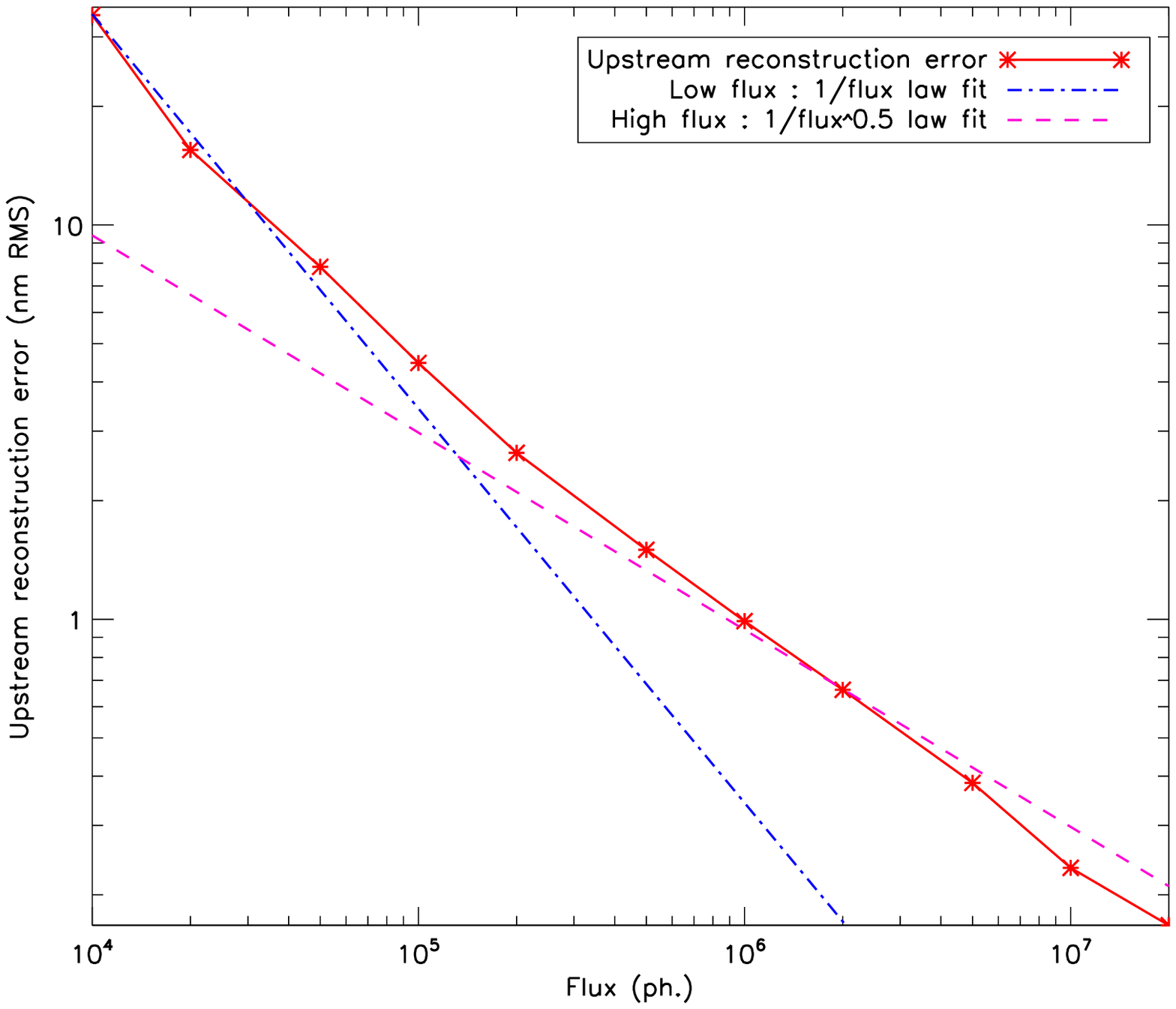}\\
\includegraphics[width = 0.8\linewidth]{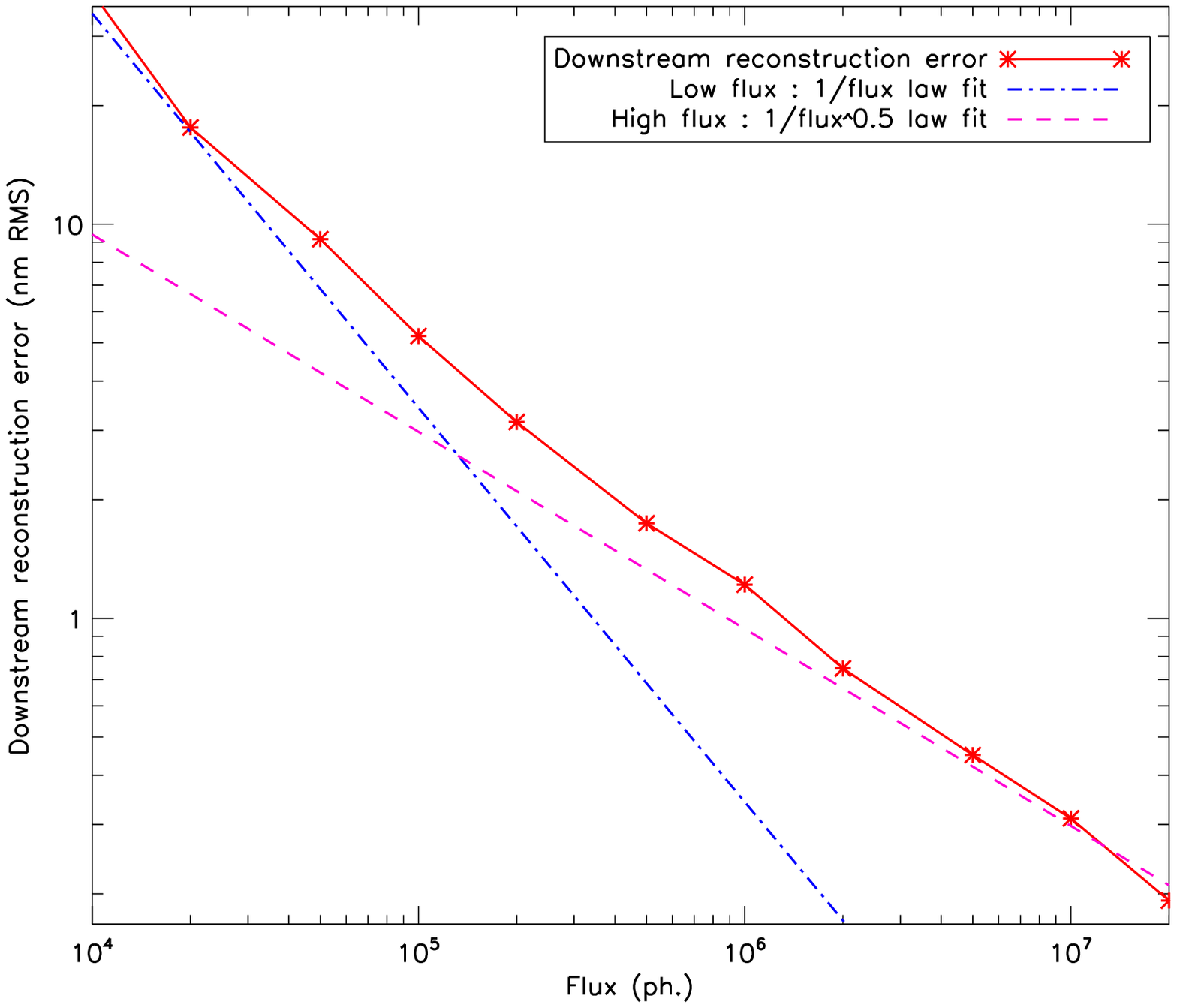}
\end{tabular}
\caption{Aberrations upstream ($\boldsymbol{\phi}_u$ (WFE $=80$
    nm), top) and downstream ($\boldsymbol{\phi}_d$ (WFE $=20$ nm), bottom) of the
  coronagraph: reconstruction error (solid red line) as a
  function of the incoming flux $\alpha$. For comparison, $\frac{1}{\alpha}$ (cyan dashed line) and
  $\frac{1}{\sqrt{\alpha}}$ (magenta dashed line) theoretical behaviours are plotted for detector
  noise only and photon noise only (respectively).}
\label{fig_SNR}
\end{figure}

The ultimate limitation of an instrument lies in the amount of noise in the
images. In Figure \ref{fig_SNR}, we present the reconstruction error for the
aberrations upstream ($\boldsymbol{\phi}_u$) and downstream
($\boldsymbol{\phi}_d$) of the coronagraph with respect to the total incoming
flux. Photon noise and detector noise ($\sigma_{\text{det}}=6\text{ e}^-$) are
added in the coronagraphic images for simulation.

The evolution of the reconstruction error presented in Figure~\ref{fig_SNR} is
proportional to ($1/\alpha$) for the detector noise limited regime (low flux)
and to ($1/\sqrt{\alpha}$) for the photon noise limited regime (high flux). In
this figure, it can be seen that for an incoming flux $\alpha\geq10^6$
photons, the reconstruction error $\epsilon_u$ for the phase upstream of the
coronagraph is smaller than $1\text{ nm RMS}$. Thus, in a calibration process,
where high values of flux ($\geq10^6$ photons)
can be easily reached, COFFEE's performance will not be significantly affected by noise.\\
It is noteworthy that the results of many similar simulations with various
levels of upstream aberrations show that COFFEE's reconstruction error does
not depend on the amplitude of the aberrations upstream of the coronagraph, as
long as the diversity phase amplitude is larger than the WFE of the
aberrations to be estimated.

\subsection{Impact of the source size on the reconstruction error}
\label{coffee_param_obj}
Our imaging model, presented in Section \ref{coffee_principe_cpd} (Eq.
\ref{eq_im_model}), assumes an unresolved object. Thus, the presence of a real
source with a given spatial extension will have an impact on the phase
reconstruction, which is quantified here. We consider here a Gaussian-shaped
laser source, emitted from a single-mode fiber. Because of the incoming light
coherence, it can be represented as a Gaussian amplitude in the entrance pupil
plane (where COFFEE assumes a uniform amplitude). Knowing this, coronagraphic
images are simulated by considering a small coherent Gaussian-shaped beam
(FWHM $\leq 0.5\frac{\lambda}{D}$) on the coronagraph, and then processed by
COFFEE.

\begin{figure}
\centering
\includegraphics[width = 0.8\linewidth]{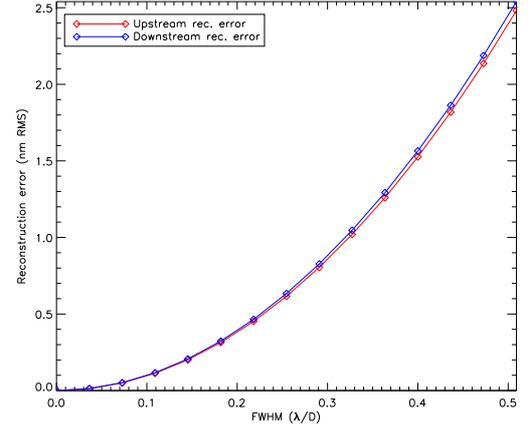}
\caption{Error reconstructions upstream (red line) and downstream (blue line)
  of the coronagraph as functions of the size of the source on the coronagraph.}
\label{fig_obj}
\end{figure}

Since the imaging model assumes an unresolved object, both reconstruction
errors for the phases upstream and downstream of the coronagraph increase with
the FWHM of the coherent object, as showed in Figure \ref{fig_obj}, but
remains low: for an FWHM smaller than $\frac{\lambda}{3D}$, the reconstruction
error is indeed sub-nanometric. The size of the laser source will thus
definitely not be a limitation for COFFEE: if this error is not negligible in
the total error budget, it is possible to include it in the imaging model used
by COFFEE (Eq. \ref{eq_im_model}) as a non-uniform (Gaussian) entrance pupil
function $\boldsymbol{P}_u(\boldsymbol{r})$.

\subsection{Sensitivity to a diversity phase error}
\label{coffee_param_errdiv}

The diversity phase
$\boldsymbol{\phi}_{div}=a_4^{div}\boldsymbol{Z}_4+a_5^{div}\boldsymbol{Z}_5$
has been defined in Section \ref{coffee_principe_cpd}. This phase
$\boldsymbol{\phi}_{div}$ is one of the inputs that COFFEE needs in order to
perform phase retrieval, so it must be calibrated as accurately as possible.
To optimize the use of COFFEE, the impact of an error on such a calibration is
studied. In this section, we consider that the diversity phase used to create
the diversity image is not perfectly known. The coronagraphic simulated
diversity image is computed with a diversity phase
$\boldsymbol{\phi}_{div}'=\boldsymbol{\phi}_{div}+\boldsymbol{\phi}_{err}$,
with $\boldsymbol{\phi}_{err}$ a randomly generated phase of given RMS value,
and COFFEE's phase reconstruction is done considering that the diversity phase
is equal to $\boldsymbol{\phi}_{div}$.
\begin{figure}
\centering
\includegraphics[width = 0.8\linewidth]{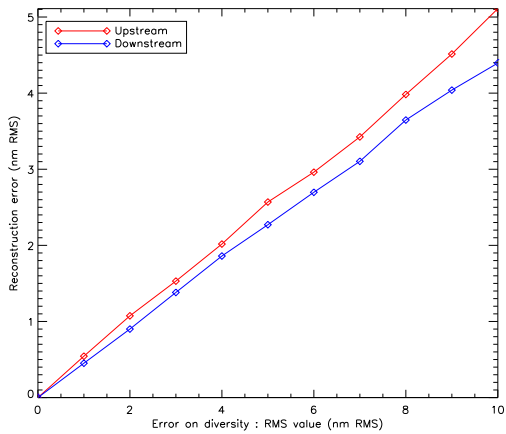}
\caption{Error reconstructions upstream (solid red line) and downstream (solid
  blue line) of the coronagraph as functions of the error on the diversity phase.}
\label{fig_errdiv}
\end{figure}
In Figure \ref{fig_errdiv}, we see that the reconstruction error increases
linearly with the calibration error on the diversity phase, with a slope of
$0.5$. Thus, the requirement on the calibration precision for the diversity
phase is typically the precision wanted for the aberration measurement.

\subsection{Impact of aliasing}
\label{coffee_param_aliasing}

The phase estimation is performed here on a truncated Zernike basis. In real
images (recorded from a bench), some speckles will originate in high-order
aberrations. These aberrations, which cannot be fitted by the truncated
Zernike basis, will have an impact on the phase estimation, called aliasing
error hereafter. Thus, it is necessary to study this aliasing error as a
function of the number of Zernike modes used in the phase reconstruction.
Here, we generate a phase on a large number of Zernike modes, and compute the
corresponding images using the perfect coronagraph model. Aberrations both
upstream and downstream of the coronagraph are then estimated by COFFEE using
an increasing number of Zernike modes. Since one of the aims of this
simulation is to determine the size of the truncated Zernike basis to be used
with experimental data recorded on an in-house bench, the noise level in the
simulated images corresponds to the one we have on this bench. The total
incoming flux is $5\ 10^6$ photons, and the detector noise is
$\sigma_{\text{det}}=1\text{ e}^-$ per pixel. Parameters used for this
simulation are gathered in Table \ref{table_aliasing}. This simulation has
been done with and without a regularization metric, so that
we can demonstrate the relevance of this metric on phase estimation.\\

\begin{figure}
\centering
\begin{tabular}{c}
\includegraphics[width = 0.8\linewidth]{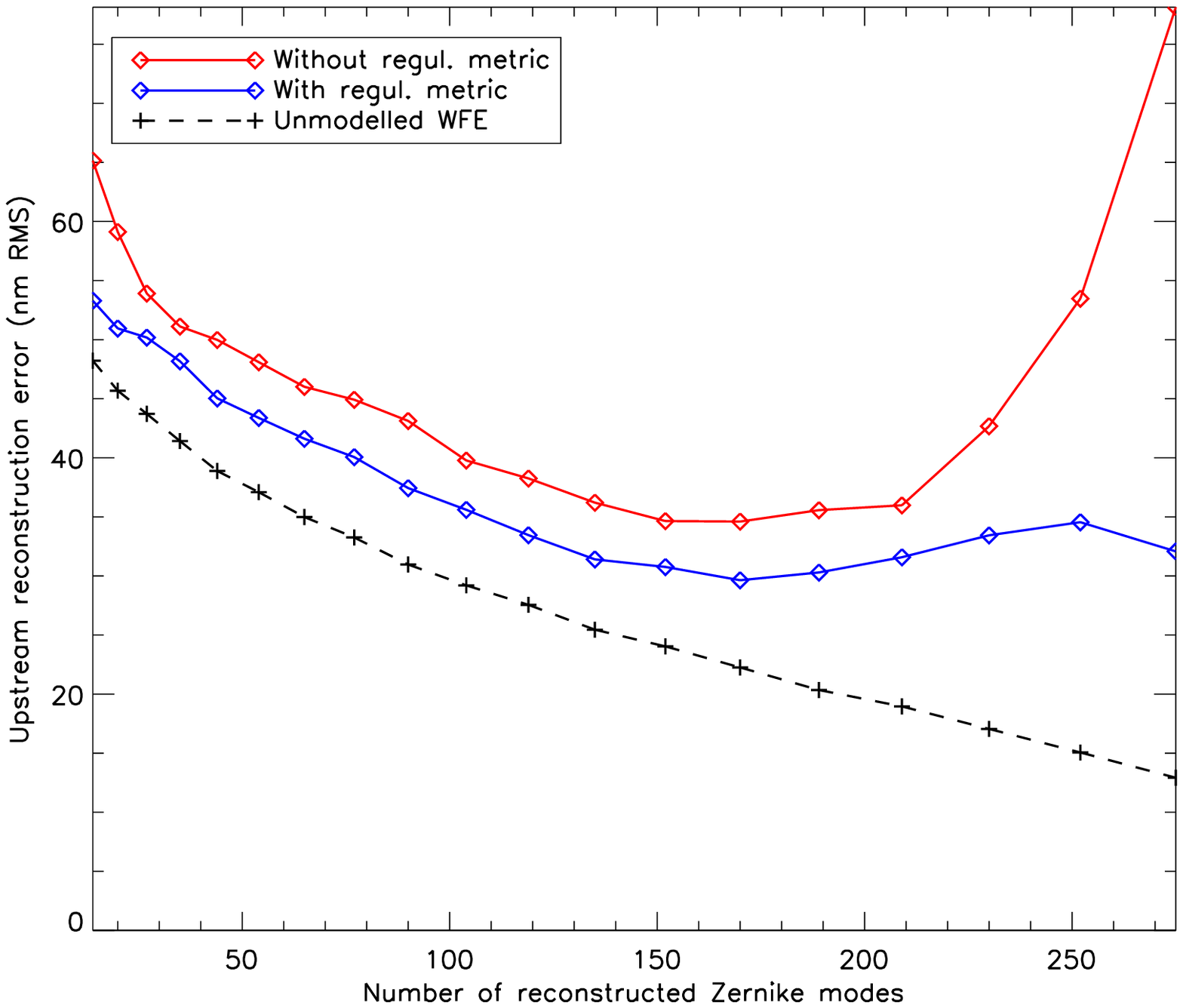}\\
\includegraphics[width = 0.8\linewidth]{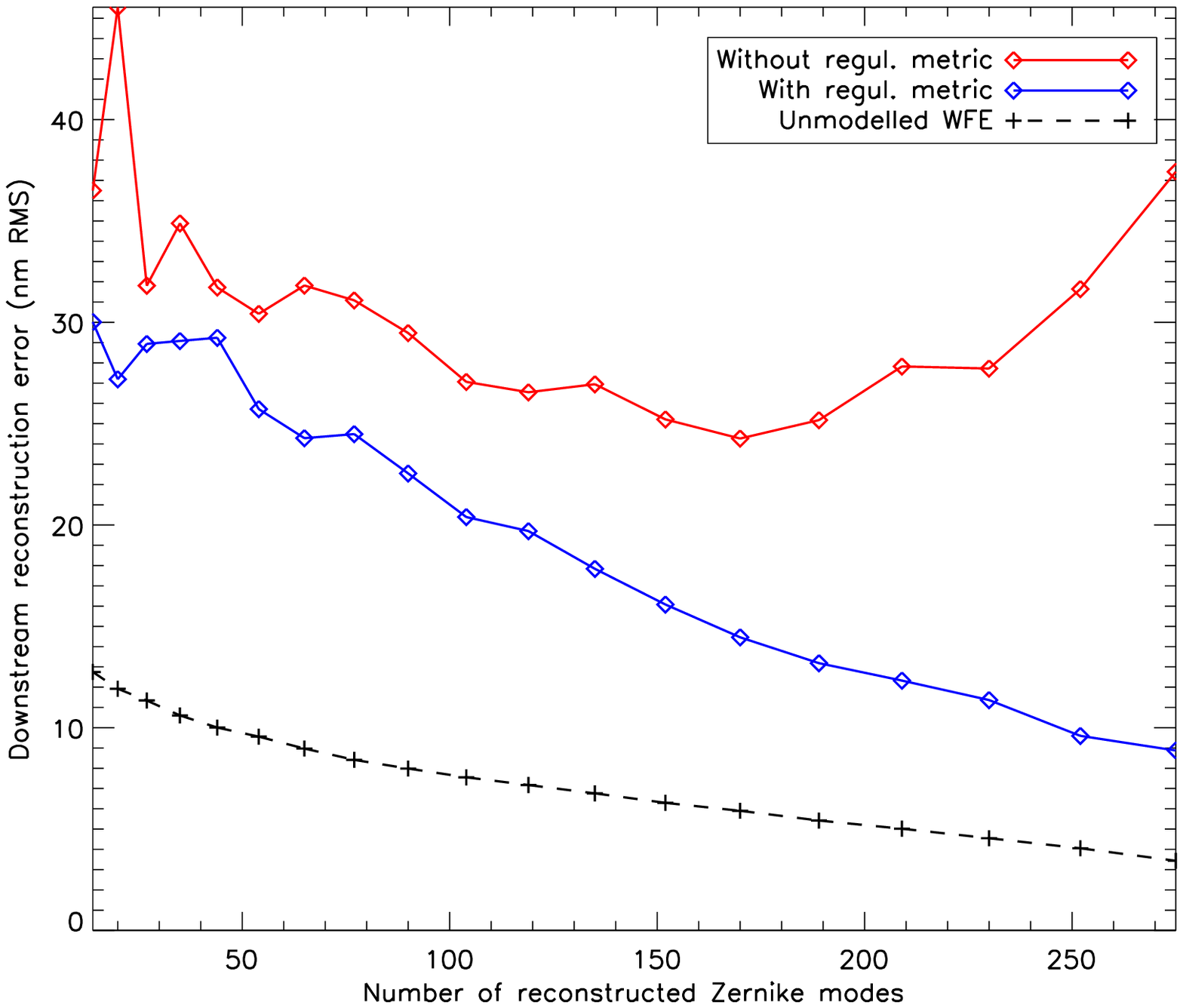}
\end{tabular}
\caption{Error reconstructions upstream (top) and downstream (bottom) of the
  coronagraph as functions of the number of reconstructed Zernike modes, with
  a regularization metric (solid blue line) and without (solid red line)}
\label{fig_aliasing}
\end{figure}

\begin{tiny}
\begin{table}
\centering
\begin{tabular}{m{4cm} m{4cm}}
  \hline
  \multicolumn{2}{c}{\textbf{Simulation}}\\
  \hline
  image size & $93\times 93$ $\frac{\lambda}{D}$ ($128\times128$ pixels, oversampling factor: $1.38$)\\
  Light spectrum & monochromatic ($\lambda=635$ nm)\\
  Aberration upstream of the coronagraph ($\boldsymbol{\phi}_u$) & WFE $=80$ nm RMS\\
  Aberration downstream of the coronagraph ($\boldsymbol{\phi}_d$) & WFE $=20$ nm RMS\\  
  Zernike basis used for $\boldsymbol{\phi}_u$ and $\boldsymbol{\phi}_d$ simulation & $350$ Zernike polynomials\\
  Incoming flux & $5\ 10^6$ photons\\
  noise & photon noise, detector noise ($\sigma_{\text{det}}=1\text{ e}^-$)\\
  \hline
  \hline
  \multicolumn{2}{c}{\textbf{COFFEE: phase estimation}}\\
  \hline
  Zernike basis used for $\boldsymbol{\phi}_u$ and $\boldsymbol{\phi}_d$ reconstruction & from $15$ to $275$ Zernike polynomials\\
  Regularization metric & With and without\\
  \hline
\end{tabular}
\caption{COFFEE: simulation parameters for studying the aliasing error.}
\label{table_aliasing}
\end{table} 
\end{tiny}
Figure \ref{fig_aliasing} presents the evolution of the reconstruction errors
when the number of reconstructed Zernike modes increases. Here, every
reconstruction error (Eq. \eqref{eq_err_rec}) is calculated on a basis of
$350$ Zernike modes; thus, the error originates both in high-order
aberrations, which are not considered by COFFEE because of the Zernike basis
finite size (modelling error), and in the impact of these high-order
aberrations on the estimated ones (aliasing). The WFE corresponding to the
aberrations that are not estimated by COFFEE (from $N$ to $350$, where N
varies between $15$ and $275$
according to Table~\ref{table_aliasing}) is called ``unmodelled WFE'' hereafter.\\
In the plot of the reconstruction error upstream of the coronagraph evolution
(Figure \ref{fig_aliasing}, top), one can see that without a regularization
metric, the reconstruction error increases for a large number of Zernike
modes. An interpretation of this behaviour is the following: because
high-order aberrations have a smaller variance, their associated speckle
intensity is lower. Thus, owing to the photon and detector noise in the image,
the SNR is smaller for these aberrations. Such behaviour leads to a trade-off
between aliasing and noise amplification for the optimal number of Zernike
modes (Figure \ref{fig_aliasing}). The best number of Zernike modes is then a
function of the aberrations level (WFE) and spectrum, as well as of the level
of noise. The use of a regularization metric allows us to avoid this noise
amplification (Figure \ref{fig_aliasing}): the reconstruction error roughly
reaches a saturation level (rather than growing to very high values).
Additionally, the use of regularization reduces the aliasing error, and avoids
the need for the difficult and somewhat \textit{ad hoc} choice
of number of Zernike modes for the reconstruction.\\
According to the results presented in Figure \ref{fig_aliasing}, we have
chosen to estimate the aberrations upstream and downstream of the coronagraph
on $170$ Zernike modes with the regularization metric of Eq. \eqref{eq_regul}.

\subsection{Model mismatch}
\label{coffee_param_model}

We have already demonstrated that ARPM images are compatible with the perfect
coronagraph model and therefore with COFFEE estimation in \cite{Sauvage-a-12}.
The Roddier \& Roddier Phase Mask (RRPM) \citep{Roddier, Roddier2} consists in
a $\pi$ phase shifting mask slightly smaller than the Airy disk. Additionally,
the use of a circular prolate function as entrance pupil apodization
$\boldsymbol{\Phi}_P$ (ARPM), proposed by \cite{corono_apod}, leads in a
perfect case (no aberrations upstream of the coronagraph) to a total
suppression of signal in the detector plane. In the simulations presented
hereafter, realistic ARPM coronagraphic images are computed following
\cite{SA_meth} to consider an accurate numerical representation of Lyot-style
coronagraphs. Then, we use COFFEE to reconstruct both phases upstream and
downstream of the coronagraph. Here, when using the formalism developed in
Section \ref{coffee_principe_im_model}, the prolate apodization function
$\boldsymbol{\Phi}_P$
is included in both simulation and reconstruction imaging models.\\

\begin{figure}
\centering
\includegraphics[width = 0.8\linewidth]{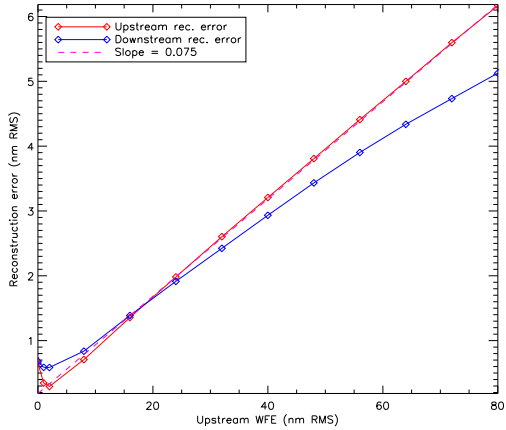}
\caption{Error reconstruction upstream of the coronagraph with respect to the
WFE of the aberration upstream of the coronagraph}
\label{fig_model}
\end{figure}

Because the perfect coronagraph model is not exactly identical to an ARPM
(although their responses to aberrations is very close), there is a model
mismatch in the estimation of aberrations upstream of the coronagraph
$\boldsymbol{\phi}_u$, which varies linearly with the WFE of
$\boldsymbol{\phi}_u$, as shown in Figure \ref{fig_model}. The model mismatch
can thus be quantified as $7.5$\% of the WFE RMS value of
$\boldsymbol{\phi}_u$, except for very small WFE ($\leq$ $1$ nm
RMS), where the variation is non-linear, but remains below $1$ nm RMS.\\
Since the variation in this model mismatch varies linearly with the WFE of
$\boldsymbol{\phi}_u$, it should not limit the ability to compensate for the
aberration upstream of an ARPM using COFFEE as focal plane wave-front
sensor (WFS).\\

\section{Laboratory demonstration}
\label{coffee_boa}
In this section we present experimental validations in the coronagraphic phase
diversity. These validations are done on the bench BOA, described in Section
\ref{coffee_boa_setup}. Section \ref{coffee_boa_cal_ab} describes a carefully
designed method developed to introduce calibrated static aberrations on the AO
bench to be measured with COFFEE. The error made on the measurements of
aberrations upstream of the coronagraph (NCPA) is quantified in Section
\ref{coffee_boa_err}. Section \ref{coffee_boa_mes} presents the static
aberration measurement performance, and Section \ref{coffee_boa_ncpa_comp}
details the procedure for compensating for the measured aberrations.

\subsection{Experimental setup}
\label{coffee_boa_setup}
\begin{figure}
\centering
\scalebox{0.4}{\input{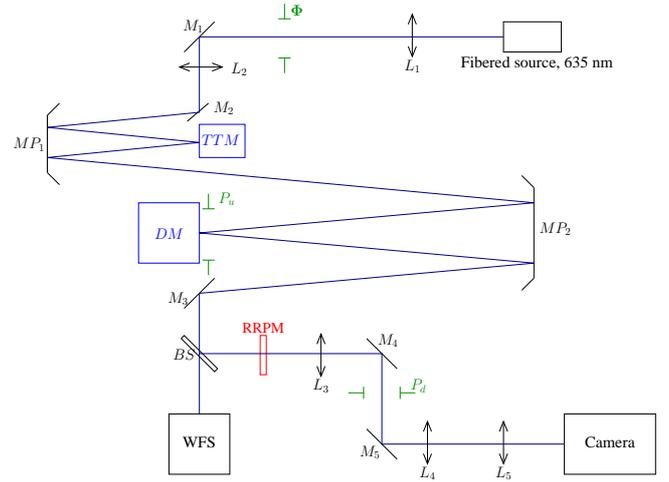}}
\caption{Adaptive optics testbed schematic representation. M$_i$: fold
  mirrors; MP$_i$: parabolic mirrors; L$_i$: lenses (doublets); BS: beam
  splitter; TTM: Tip-Tilt mirror; DM: Deformable mirror; RRPM: coronagraphic
  focal plane mask; $\boldsymbol{\Phi}$: prolate apodizer; WFS: AO wave-front
  sensor}
\label{boa}
\end{figure}

Figure \ref{boa} shows the design of our in-house bench. The input beam,
emitted from a fibered laser source ($\lambda = 635$ nm) comes through the
prolate apodizer $\boldsymbol{\Phi}$, which is in the entrance pupil plane
($P_u$). The beam is reflected by the tip-tilt mirror (TTM) and then by the
deformable mirror (DM, entrance pupil, $D_u=40$ mm, $6\times 6$ actuators).
The beam-splitter sends a fraction of the beam to the AO wave-front sensor
(Shack-Hartmann, $5\times 5$ sub-apertures). On the other channel, the light
is focused onto a RRPM, whose diameter is $d_c=18.1\ \mu$m (angular diameter
is $1.06\frac{\lambda}{D_u}$). After going through the Lyot stop plane ($P_d$,
with $D_d=0.99D_u$), the beam is focused onto the camera ($256\times 256$
pixels images with an oversampling of $2.75$, detector noise
$\sigma_{\text{det}}=1\text{ e}^-$). For faster computations, recorded images
are re-binned to $128\times 128$ pixels images with an oversampling of $1.38$.

\subsection{Introduction of calibrated aberrations}
\label{coffee_boa_cal_ab}

\begin{figure}
  \centering
\begin{tabular}{cc}
\includegraphics[width = 0.25\linewidth]{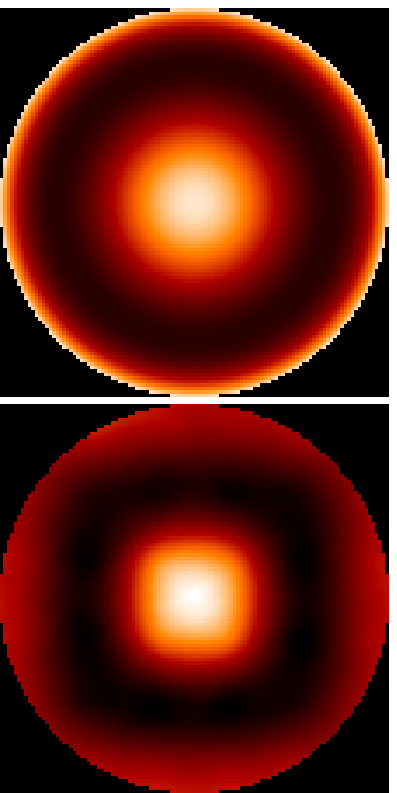}&
\includegraphics[width = 0.60\linewidth]{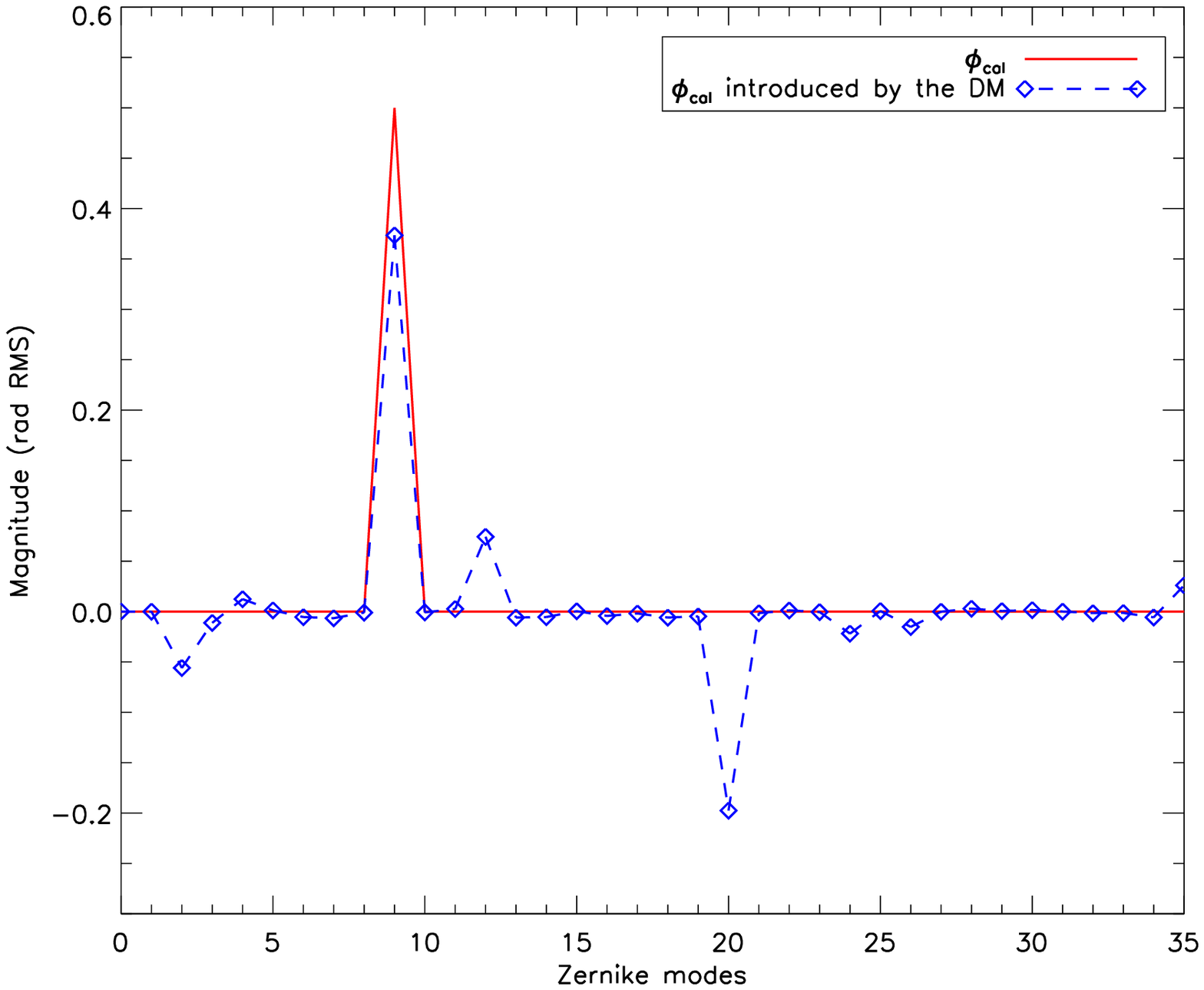}\\
\end{tabular}
\caption{Introduction of calibrated aberration on BOA: case of a pure
  spherical aberration. Left: theoretical wave-front (top) and DM introduced
  wave-front (bottom). Right: corresponding Zernike modes for the theoretical
  introduced aberration (solid red line) and the DM introduced aberration
  (dashed blue line).}
\label{boa_ab_sph}
\end{figure}

To evaluate COFFEE's performance, we introduce calibrated aberrations
on the bench using a process described in this section. We consider an
aberration phase $\boldsymbol{\phi}_{cal}$ to be introduced on BOA. First,
since the phase is represented by the DM with a finite number of actuators
($6\times 6$), the introduced aberration will not match the
aberration $\boldsymbol{\phi}_{cal}$ perfectly, as illustrated in Figure
\ref{boa_ab_sph} in the case of a pure spherical aberration.
  
Our aim is here to introduce, using the DM, the closest aberration to the
aberration $\boldsymbol{\phi}_{cal}$. We let $\boldsymbol{F}$ be the DM
influence matrix (obtained by calibration); any DM introduced aberration
$\boldsymbol{\phi}^{DM}$ can be described as a set of actuator voltages
$\boldsymbol{u}$ ($\boldsymbol{\phi}^{DM}=\boldsymbol{F}\boldsymbol{u}$). We
are thus looking for the set $\boldsymbol{u}_{cal}$ which solves the
least-squares problem:
\begin{equation}
\boldsymbol{u}_{cal}=\underset{\boldsymbol{u}}{\arg\min}\left\| \boldsymbol{F}\boldsymbol{u} - \boldsymbol{\phi}_{cal} \right\|^2.
\end{equation}
The solution of this problem can be written as
\begin{equation}
\boldsymbol{u}_{cal}=\boldsymbol{T}\boldsymbol{\phi}_{cal}\text{,}
\end{equation}
with $\boldsymbol{T}$ the generalized inverse of matrix $\boldsymbol{F}$.
Using the interaction matrix $\boldsymbol{D}$ (resulting from calibration), we
can compute the corresponding set of slopes $\boldsymbol{s}_{cal}$
($\boldsymbol{s}_{cal}=\boldsymbol{D}\boldsymbol{u}_{cal}$), which can then be
used to modify the AO loop reference slopes $\boldsymbol{s}_{\text{ref}}$.
Thus, closing the AO loop with the reference slopes
$\boldsymbol{s}_{\text{ref}}+\boldsymbol{s}_{cal}$, we introduce an aberration
$\boldsymbol{\phi}^{DM}_{cal} =
\boldsymbol{F}\boldsymbol{u}_{cal}=\boldsymbol{F}\boldsymbol{T}\boldsymbol{\phi}_{cal}$
on the bench,
which is the best fit of $\boldsymbol{\phi}_{cal}$ in the least squares sense.\\
We also have to consider that the bench BOA presents its own unknown static
aberrations $\boldsymbol{\phi}^{\text{BOA}}_u$ and
$\boldsymbol{\phi}^{\text{BOA}}_d$ upstream and downstream of the coronagraph
(respectively). Thus, if a calibrated aberration $\boldsymbol{\phi}_{cal}$ is
introduced in the entrance pupil, aberrations $\boldsymbol{\phi}_u$ upstream
of the coronagraph will be
\begin{equation}\label{eq_ncpa_boa}
\boldsymbol{\phi}_u=\boldsymbol{\phi}_{cal}+\boldsymbol{\phi}^{\text{BOA}}_u.
\end{equation}
To get rid of the unknown aberration $\boldsymbol{\phi}^{\text{BOA}}_u$, we
perform a differential phase estimation: 
\begin{enumerate}
\setlength\itemsep{-0.1in}
\item We introduce the aberration $\boldsymbol{\phi}^{DM}_{cal}$
  on the bench. A phase $\hat{\boldsymbol{\phi}}_{u}^+=\hat{\boldsymbol{\phi}}^{DM}_{cal}+\hat{\boldsymbol{\phi}}^{BOA}_u$
  is estimated using
  focused and diverse images recorded on the camera.\\
\item The opposite aberration $-\boldsymbol{\phi}^{DM}_{cal}$ is then
  introduced. A phase
  $\hat{\boldsymbol{\phi}}_{u}^-=-\hat{\boldsymbol{\phi}}^{DM}_{cal}+\hat{\boldsymbol{\phi}}^{BOA}_u$ is estimated.\\
\item  The half difference
  $\hat{\boldsymbol{\phi}}^{DM}_{cal}=\frac{\hat{\boldsymbol{\phi}}_{u}^+-\hat{\boldsymbol{\phi}}_{u}^-}{2}$ is our
  estimate of $\boldsymbol{\phi}_{cal}$.
\end{enumerate}

The first use of this process is to calibrate the diversity phase itself.
Since this phase will be introduced using the AO system, the actually
introduced diversity phase will not exactly match the theoretical mix of
defocus and astigmatism. We introduce the aberrations
$\boldsymbol{\phi}_{div}$ and $-\boldsymbol{\phi}_{div}$ on the bench using
the AO system. These two aberrations are then estimated using classical phase
diversity (no coronagraph), with a pure defocus of diversity phase introduced
using a flat
glass plate of known thickness $e$ in a focused beam.\\
Such a process gives us an accurate estimation of the diversity phase really
introduced on the bench, with an estimated accuracy of $4$ nm RMS on the
introduced aberration. This calibration is then used in COFFEE's estimations
performed on experimental images.

\subsection{Performance assessment: error budget}
\label{coffee_boa_err}

From simulations presented in Section \ref{coffee_param}, we establish an
error budget for estimating aberrations upstream of the coronagraph using
experimental data:

\begin{itemize}
\renewcommand{\labelitemi}{$\diamond$}
\setlength\itemsep{-0.1in}
\item Photon and detector noise error: on the BOA bench, the typical
  incoming flux is $f_{\text{BOA}}=5\ 10^6$ photons. Knowing that we have
  photon noise and a detector noise with $\sigma_{\text{det}}=1\text{
    e}^-$, we can evaluate the noise error: $\epsilon_{\text{noise}}=0.9$ nm RMS.\\
\item The diversity phase $\boldsymbol{\phi}_{div}$ has been calibrated using
  classical phase diversity, using the process presented in Section
  \ref{coffee_boa_cal_ab}. Such an estimation has been performed with an error
  of $4.0$ nm RMS (value calculated from an error budget evaluated for a
  classical phase diversity estimation on the BOA bench. Such accuracy has
  already been obtained on this bench by \cite{Sauvage-a-07}). According to
  Section \ref{coffee_param_errdiv}, this error on the diversity phase leads
  to an error
  $\epsilon_{\text{model}}= 2.0$ nm RMS.\\
\item The source is a coherent Gaussian-shaped beam whose FWHM is
  $0.27\frac{\lambda}{D}$ on the coronagraph. According to the simulations of
  Section \ref{coffee_param_obj}, this leads to a
  reconstruction error: $\epsilon_{\text{obj}}=0.7$ nm RMS.\\
\item Residual turbulent speckles, which originate in uncorrected turbulent
  aberrations, are not included in the imaging model. To measure the impact of
  these speckle on the reconstruction, several wave-fronts have been
  successively recorded using a commercial Shack-Hartmann wave-front sensor.
  From these acquisitions, we calculate the WFE of the residual turbulent
  phase: $\sigma_{\phi_{\text{turb}}}=1.2$ nm RMS. This residual turbulence
  will create speckles on the detector, which will be considered by COFFEE as
  originating in NCPA. Thus, the residual turbulence error
  $\epsilon_{\text{turb}}$ made by COFFEE is estimated to
  $\epsilon_{\text{turb}}=\sigma_{\phi_{\text{turb}}}=1.2$ nm RMS.\\
\item Aliasing error, which originates in high-order aberrations, has
  been studied in Section \ref{coffee_param_aliasing}. For a phase upstream of
  the coronagraph estimated on $N=170$ Zernike modes, we have
  $\epsilon_{\text{aliasing}}=18.3$ nm RMS.\\
\item From simulations, we know that the model mismatch is $7.5$\% of WFE. For
  this study, we will not estimate aberrations with a WFE stronger than $80$
  nm RMS. For such a WFE, the model error is $\epsilon_{\text{model}}=6.0$ nm
  RMS.
\end{itemize}
\begin{table}
\centering
\begin{tabular}{m{4.0cm} m{4.0cm}}
  \hline
  \multicolumn{2}{c}{\textbf{Error budget}}\\
  \hline
  Noise& $\epsilon_{\text{noise}}= 0.9$ nm RMS\\
  Model mismatch& $\epsilon_{\text{model}}= 6.0$ nm RMS\\
  Error on diversity& $\epsilon_{\text{div}}=2.0$ nm RMS\\
  Resolved object& $\epsilon_{\text{obj}}=0.7$ nm RMS\\
  Residual turbulence& $\epsilon_{\text{turb}}=1.2$ nm RMS\\
  Aliasing& $\epsilon_{\text{aliasing}}=18.3$ nm RMS\\
  \hline
  \textbf{Total error}& $\epsilon=\sqrt{\sum_{i}\epsilon_i^2}=20.6$ nm
  RMS\\
  \textbf{Total error per Zernike mode}& $\epsilon'= 1.6$ nm RMS per estimated Zernike mode\\
  \hline
\end{tabular}
\caption{COFFEE: error budget for the estimation of an aberration upstream of
  the coronagraph on BOA.}
\label{error_budget}
\end{table}

As one can see in Table \ref{error_budget}, the error budget is mainly driven
by the aliasing error. The second most important term is the model mismatch
(even though it goes to zero with the WFE).

\subsection{Measurement of aberrations upstream of the coronagraph}
\label{coffee_boa_mes}

In this section, we introduce calibrated aberrations on the BOA bench upstream
of the coronagraph, and then estimate them with COFFEE in order to evaluate
its performance. In the course of this study, we realized that the position of
the coronagraphic image on the detector (quantified by the tip-tilt downstream
of the coronagraph) is a critical issue. Indeed, it occurred that COFFEE was
able to perform phase retrieval only for downstream tip-tilt $[a_2,a_3]$
values within the range $[-100\text{ nm RMS}; 100\text{ nm RMS}]$
($[-\frac{\lambda}{6D}; \frac{\lambda}{6D}]$). To get rid of this constraint,
we have developed a method to perform a preliminary estimation of the tip-tilt
downstream of the coronagraph. This method, which uses the diversity image, is
fully described in Appendix \ref{tt_down}.

\subsubsection{Measurement of tip-tilt upstream of the coronagraph}
\label{coffee_boa_tt}

We present the estimation of a tilt aberration upstream of the coronagraph
using COFFEE in this section. Using the AO system, we introduce a tilt
aberration by adding a constant value $\delta \boldsymbol{s}_{\text{TT}}$ to
the AO wave-front sensor references slopes $s_{\text{ref}}$, and then closing
the AO loop on the slopes $\boldsymbol{s}_{\text{ref}}+\delta
\boldsymbol{s}_{\text{TT}}$. To accurately calibrate the introduced tilt, for
each position, we first estimate the aberrations using classical phase
diversity (no coronagraph). Then, the RRPM is put in the focal plane, and the
same operation is repeated: for each position, we record two images, and then
estimate the aberrations using COFFEE.

\begin{figure}
\centering
\includegraphics[width = 0.8\linewidth]{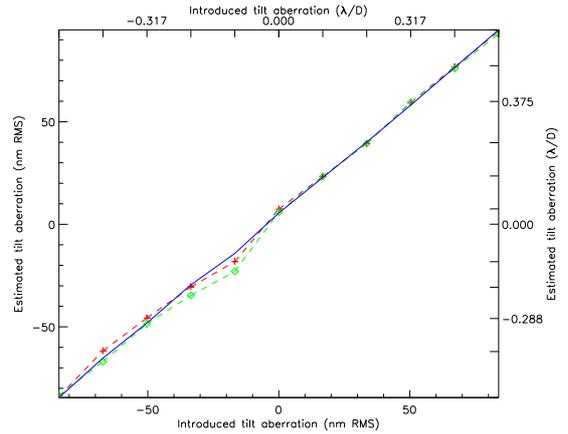}
\caption{Estimation of a tilt aberration on BOA: calibration (solid blue line)
  and COFFEE's estimation with bound on the tip-tilt downstream of the
  coronagraph (dashed crossed red line) and without boundaries (dashed diamond
  green line) }
\label{boa_tt}
\end{figure}
 
From the upstream tilt reconstruction performed by COFFEE (Figure
\ref{boa_tt}), we calculate an average reconstruction error:
$\epsilon_{\text{tilt}}=2.1$ nm. Part of this error is due to an error on the
estimation of tip-tilt downstream of the coronagraph. An improved estimation
has been performed by setting boundaries on the downstream tip-tilt. Its value
is evaluated before COFFEE's estimation using the method described in appendix
\ref{tt_down} with the diversity coronagraphic image recorded for a tip-tilt
upstream the coronagraph value close to $0$ nm RMS (centered coronagraph).
Such an estimation process gives us an estimation of tip-tilt downstream of
the coronagraph $\{a_2^{\text{do}},a_3^{\text{do}}\}$ with an accuracy of $\pm
1.5$ nm RMS. Using this estimation as the starting value for the minimization,
and setting bounds of $\pm 1.5$ nm RMS on it, we processed the same
experimental data. This, in turn, results in a better estimation of tilt
upstream of the coronagraph (Figure \ref{boa_tt}), with an average error
$\epsilon_{\text{tilt}}=1.5$ nm, which is close to the expected error per
Zernike mode given in Section \ref{coffee_boa_err} ($\epsilon'= 1.6$ nm RMS).

\subsubsection{NCPA measurements}
\label{coffee_boa_ncpa_mes}

In this section, we introduce aberrations upstream of the coronagraph. The
aberration $\boldsymbol{\phi}_{cal}$ is expanded on the first $15$ Zernike
modes (which is the largest number of modes we can properly describe with our
$6 \times 6\ DM$), and then we estimate these aberrations using COFFEE,
following the process described in Section \ref{coffee_boa_cal_ab}. To take
the DM action into account on the introduced phase (illustrated in Figure
\ref{boa_ab_sph}), aberrations $\boldsymbol{\phi}_{cal}$ are first estimated
with classical phase diversity (no phase mask in the coronagraphic focal plane
\citep{Sauvage-a-07}). This estimation gives us a calibration of the
introduced aberration, which is then used to evaluate the accuracy of COFFEE's
estimation.

\begin{figure}
\centering
\begin{tabular}{c}
\includegraphics[width =0.8\linewidth]{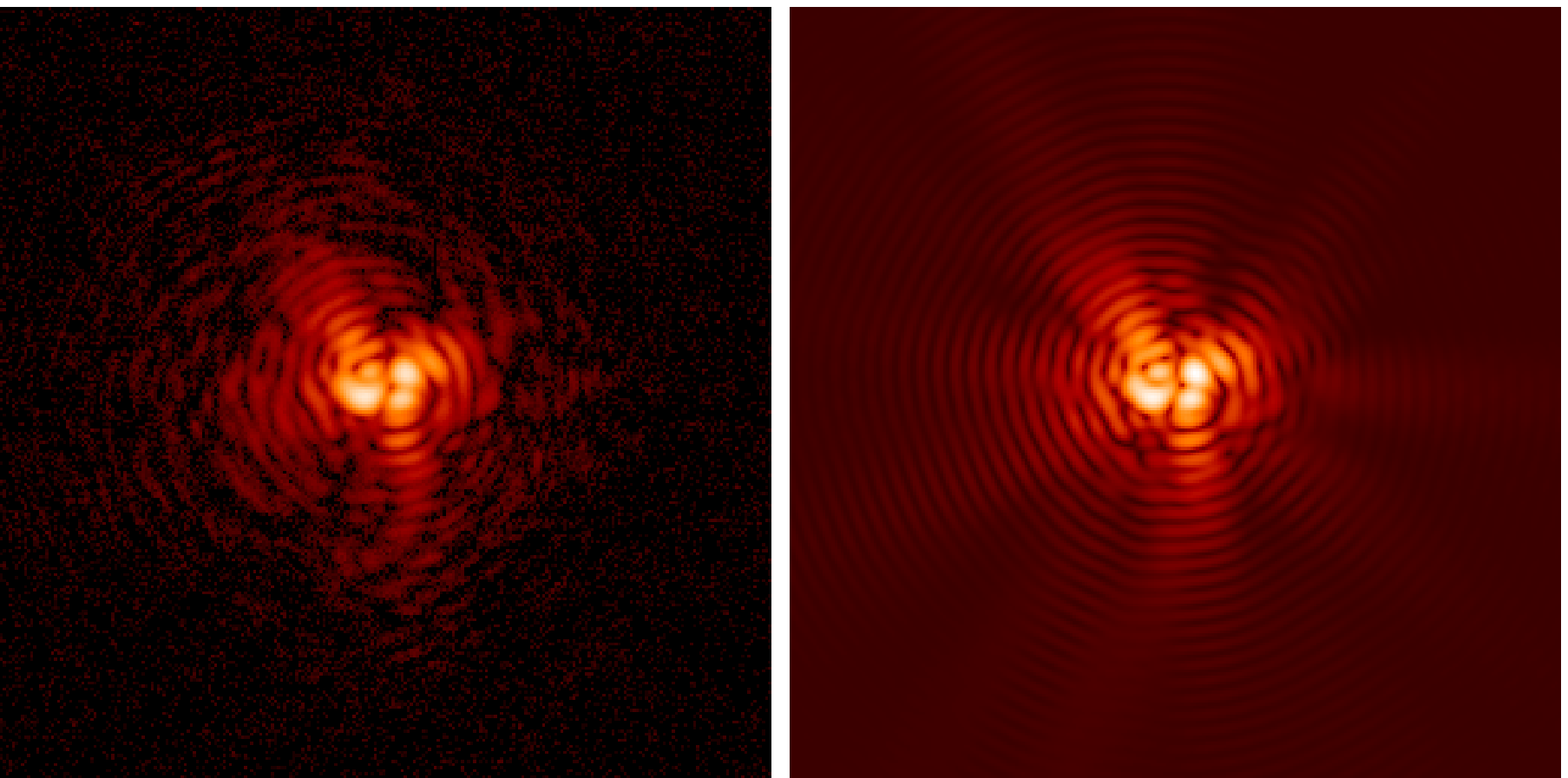}\\
\includegraphics[width =0.8\linewidth]{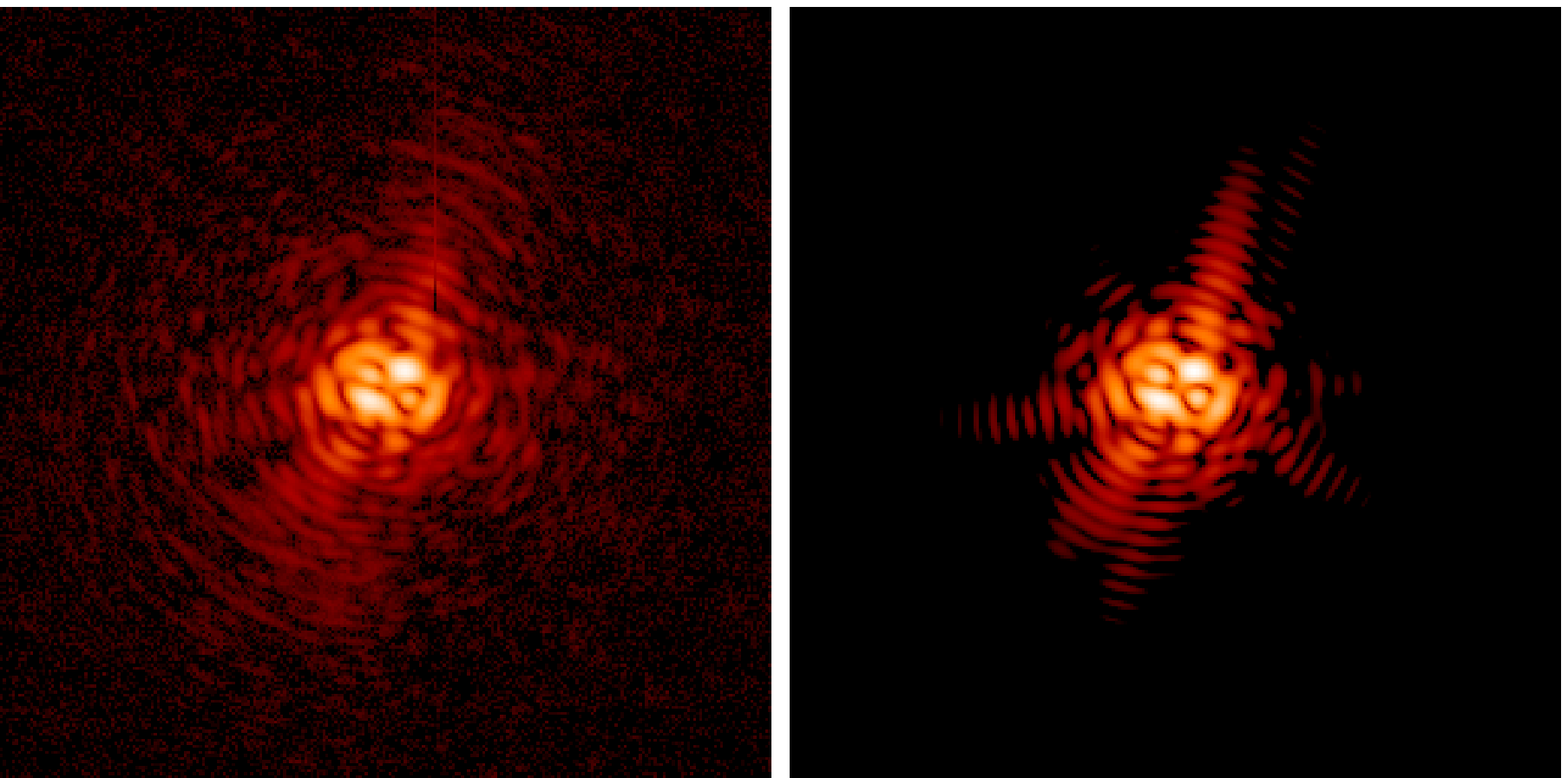}\\
\includegraphics[width = 0.8\linewidth]{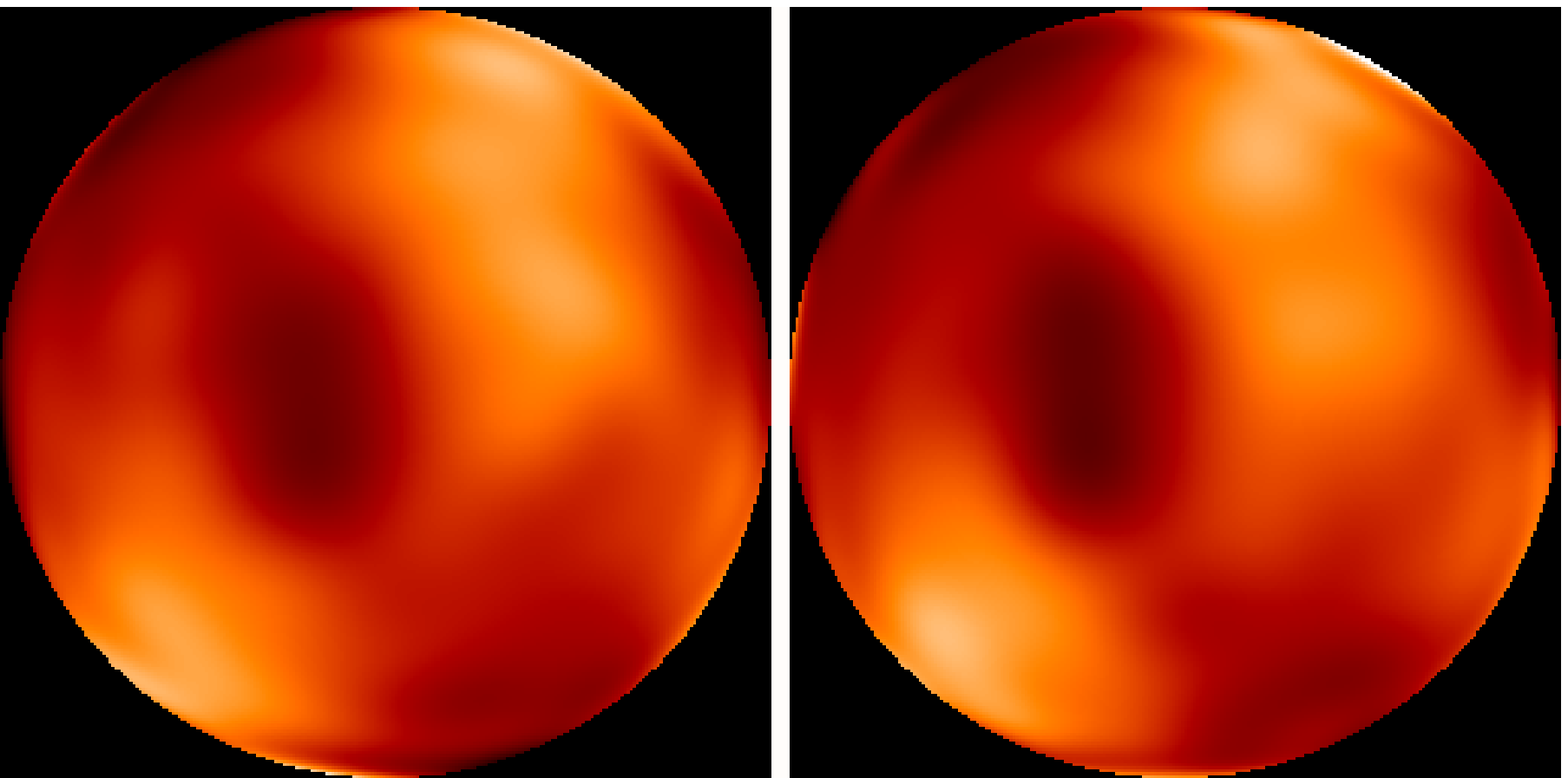}
\end{tabular}
\caption{COFFEE: NCPA estimation of an introduced phase
  $\boldsymbol{\phi}_{cal}$ on BOA. Top: for an aberration $+\boldsymbol{\phi}_{cal}$,
  recorded coronagraphic image from the bench (left) and computed image using
  the reconstructed aberration $\hat{\boldsymbol{\phi}}_u^+$ (right) (log. scale, same
  range for both images). Middle: same images
  for an aberration $-\boldsymbol{\phi}_{cal}$ introduced and a reconstructed aberration
  $\hat{\boldsymbol{\phi}}_u^-$ (log. scale, same
  range for both images). Bottom: calibrated introduced aberration (left) and
  COFFEE estimated aberration (right)}
\label{ncpa_mes}
\end{figure}

At convergence of the reconstruction, a very good match can be observed
between the experimental images and the ones computed for the estimated
aberrations (Figure \ref{ncpa_mes}, top and middle). This, in turn, results in
a very good match between the aberrations measured by COFFEE (Figure
\ref{ncpa_mes},
right) and the introduced ones (Figure \ref{ncpa_mes}, left).\\
From the experimental phase estimation presented in Figure \ref{ncpa_mes}, we
compute a reconstruction error between the classical diversity phase
calibrated aberration and COFFEE's estimation:
\begin{equation}
\epsilon_{\text{exp}}=22.5\ \text{nm RMS}.
\end{equation}
One can notice that this error is close to the expected error budget, i.e.
that there is a good match between the performance assessment study
carried out in Section \ref{coffee_param} and the experimental results
presented in this section. 

\subsection{Low-order NCPA compensation}
\label{coffee_boa_ncpa_comp}

Lastly, the ability of COFFEE to compensate for the aberrations upstream of
the coronagraph is experimented on BOA. In Section \ref{coffee_boa_mes}, the
aberrations upstream of the coronagraph are expanded on $170$ Zernike modes,
in order to have the smallest
reconstruction error (according to Section \ref{coffee_param_aliasing}).\\
As previously mentioned, the compensation on BOA is limited to the
$15^{\text{th}}$ Zernike mode. Thus, what is required in a closed loop process
is the most accurate estimation of $15$ Zernike modes rather than an accurate
measurement of every estimated Zernike mode. Using a basis of $36$ Zernike
modes for the reconstruction is sufficient to give an accurate estimation of
the first $15$ Zernike modes: the aliasing error, which is the most important
error source, will mainly degrade the estimation accuracy of the
reconstructed high orders (close to $Z_{36}$).\\
To demonstrate the ability of COFFEE to be used in a closed loop, we introduce
a set of aberrations on the $DM$ by modifying the reference slopes, as
described in Section \ref{coffee_boa_cal_ab}. Then, we use the pseudo-closed
loop (PCL) method described in \cite{Sauvage-a-07}. This iterative process has
two stages: for the PCL iteration i:
\begin{enumerate}
\setlength\itemsep{-0.1in}
\item acquisition of the focused $\boldsymbol{i}_{c}^{f}$ and diverse
    $\boldsymbol{i}_{d}^{f}$ images;\\
\item estimation of the aberration $\hat{\boldsymbol{\phi}}_u^i$ upstream of
  the coronagraph;\\
\item computation of the corresponding reference slopes correction $\delta
  \boldsymbol{s}=g\boldsymbol{DT}\hat{\boldsymbol{\phi}}_u^i$, where
  $\boldsymbol{D}$ and $\boldsymbol{T}$ are the interaction and influence
  matrices defined in Section \ref{coffee_boa_cal_ab} and
  $g$ is the PCL gain;\\
\item the AO loop is closed on the modified reference slopes.
\end{enumerate}

The computation time (step 2) varies from $1$ minute to $2.5$ minutes,
allowing us to compensate for quasi-static aberrations upstream of the
coronagraph. This compensation process is limited by the estimation accuracy
of the first $15$ Zernike modes performed by COFFEE, which corresponds to the
error budget established in Section \ref{coffee_boa_err}), and by the ability
of the DM to reproduce a given wave-front. Indeed, the correction introduced
on the bench (step $2$ of the PCL process) is the best fit of the estimated
phase $\hat{\boldsymbol{\phi}}_u^i$ in the least-square sense (as presented in
Section \ref{coffee_boa_cal_ab}). The difference between the estimated
aberration and the actual introduced correction will thus limit the
compensation performance of the PCL process. Considering these two
limitations, one can compute the variance $\sigma_{\text{BOA}}^2$ (for the
first $15$ Zernike modes) that can be reached on the BOA bench:
\begin{equation}
\sigma_{\text{BOA}}^2=4.4\ 10^{-2}\ \text{rad RMS}^2.
\end{equation} 

\begin{figure}
\centering
\includegraphics[width =0.8\linewidth]{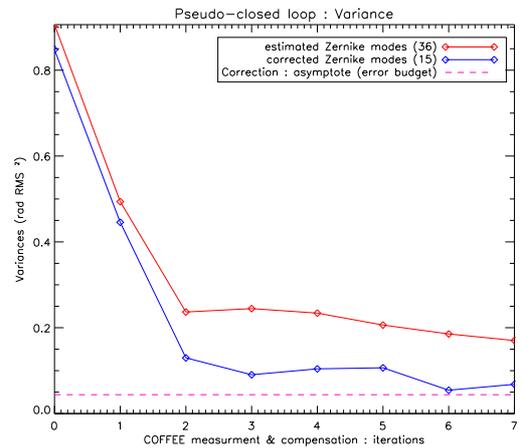}
\caption{PCL on the bench BOA ($g_{\text{PCL}}=0.5$): variance of the residual
  static aberrations upstream of the coronagraph for the $36$ COFFEE estimated
  Zernike modes (solid red line) and the $15$ corrected modes (solid blue
  line). The magenta dashed line represents the ultimate performance one can
  reach according to the error budget detailed in \ref{coffee_boa_err}}
\label{ncpa_comp}
\end{figure}

The correction and stabilization of the NCPA variance can be seen in Figure
\ref{ncpa_comp}. One can see that the variance of the $15$ corrected Zernike
modes reaches the expected asymptotic value $\sigma_{\text{BOA}}^2$. This
result is the very first demonstration of COFFEE's ability to compensate for
aberrations upstream of the coronagraph. A compensation at levels compatible
with SPHERE or GPI-like instruments will require th using a DM with many more
actuators, and working on the reduction of the dominant term of the error
budget, which is aliasing.

\section{Conclusion}
\label{ccl}

In this paper, we have presented a thorough simulation study (Section
\ref{coffee_param}) and a first experimental validation (Section
\ref{coffee_boa}) of the coronagraphic wave-front sensor called COFFEE, which
consists mainly in the extension of the phase diversity concept to a
coronagraphic imaging system. From the validation and careful performance
assessment of COFFEE, we showed that COFFEE is currently limited by the
aliasing error, due to high-order
aberrations, which are difficult to model with a Zernike basis.\\
In Section \ref{coffee_boa}, we presented a first experimental validation of
COFFEE using an ARPM. We introduced calibrated aberrations upstream of the
coronagraph (NCPA), using the AO sub-system, and estimated them with COFFEE.
The accuracy we obtained on these estimation shows a very good match with our
error budget. Lastly, we used COFFEE in an iterative process to perform a
preliminary validation of COFFEE's ability to
compensate for the aberrations upstream of the coronagraph.\\
Several perspectives are currently considered to optimize COFFEE: firstly, in
order to minimize the impact of the aliasing error on the phase
reconstruction, we plan to perform the phase reconstruction on a pixel-wise
map, which is more suitable than a truncated Zernike basis. Secondly, we would
like to improve the imaging model, both to make COFFEE work with other
coronagraph than the ARPM and to reduce the model error, which is currently
the second most important one, even though it goes to zero with the WFE. Two
solutions are considered. In the absence of residual turbulence, an accurate
imaging model is obtained by propagating the electric field through each plane
of the coronagraphic imaging system (Figure \ref{syst_coro}) for an arbitrary
focal plane coronagraphic mask. Such a method, where no model error needs to
be considered, can be used for a laboratory calibration. Alternatively, a more
accurate analytical imaging model, which could include a residual turbulent
aberration, can be developed. Such a model, which could include a residual
turbulent aberration, will ultimately allow us to perform NCPA estimation on
images from the sky. These improvements should allow us to estimate and
compensate for the aberrations upstream of the coronagraph using COFFEE with a
nanometric precision in a closed loop
process.\\
A further perspective is to extend COFFEE to phase and amplitude aberration
estimation, in order to create a dark hole region in the coronagraphic image.

\begin{acknowledgements}
  The authors would like to thank Mamadou N'Diaye, Kjetil Dohlen and Thierry
  Fusco for stimulating discussions, as well as Marc Ferrari, David Mouillet
  and Jean-Luc Beuzit for their support, and the R\'egion
  Provence-Alpes-C\^ote d'Azur for partial financial support of B. Paul's
  scholarship. This work has been partially funded by the European Commission
  under FP7 Grant Agreement No. 312430 Optical Infrared Coordination Network
  for Astronomy.
\end{acknowledgements}

\appendix
\section{Implementation details}
\label{impl_details}

COFFEE performs a phase estimation by minimizing a criterion $J$ whose
expression is given by equation \ref{eq-pb-inverse}. To estimate
$\boldsymbol{\phi}_u$ and $\boldsymbol{\phi}_d$ (expanded on a truncated
Zernike basis), we need both gradients $\frac{\partial
  J}{\partial\boldsymbol{a}_u}$ and $\frac{\partial
  J}{\partial\boldsymbol{a}_d}$, where
$\boldsymbol{a}_x=\{a_{x_1},a_{x_2},...,a_{x_N}\}$ is a vector that contains
the Zernike coefficients, for an aberration expanded on $N$ Zernike modes ($x$
is
for $u$ (upstream) or $d$ (downstream)).\\
Let us write the numerical expression of $J^{\text{foc}}$, using the notations
defined in Section \ref{coffee_principe_cpd}:
\begin{equation}
\begin{aligned}
J&=\frac{1}{2}\sum_{t} \left |\frac{\boldsymbol{i}_c^{\text{foc}}[t] 
-\alpha.\boldsymbol{h}_{\text{det}}[t]\star\boldsymbol{h}_c^{\text{foc}}[t]-\beta}{\boldsymbol{\sigma}_n^{\text{foc}}[t]}
\right |^2\\
&+\frac{1}{2}\sum_{t} \left |\frac{\boldsymbol{i}_c^{\text{div}}[t] 
-
\alpha.\boldsymbol{h}_{\text{det}}[t]\star\boldsymbol{h}_c^{\text{div}}[t]-\beta}{\boldsymbol{\sigma}_n^{\text{div}}[t]}
\right |^2\\
&+\mathcal{R}_{\boldsymbol{\phi}_u}+\mathcal{R}_{\boldsymbol{\phi}_d}\\
& = J^{\text{foc}}+J^{\text{div}}+\mathcal{R}_{\boldsymbol{\phi}_u}+\mathcal{R}_{\boldsymbol{\phi}_d}.
\end{aligned}
\end{equation}
With $t$ the pixel position in the detector plane.
$\boldsymbol{\sigma}_n^{\text{foc}}$ and $\boldsymbol{\sigma}_n^{\text{div}}$
are the noise variance maps. Considering the expression of $J$, we derive
$J^{\text{foc}}$, and then deduce the gradients expressions of
$J^{\text{div}}$ using a trivial substitution. Expressions of the
regularization terms gradients $\frac{\partial
  \mathcal{R}_{\boldsymbol{\phi}_x}}{\partial \boldsymbol{a}_x}$ are given by
\begin{equation}
\frac{\partial \mathcal{R}_{\boldsymbol{\phi}_x}}{\partial \boldsymbol{a}_x}=R_{\boldsymbol{a}_x}^{-1}\boldsymbol{a}_x.
\end{equation}  
The calculation  of gradients $\frac{\partial J}{\partial\boldsymbol{\phi}_u}$ and
$\frac{\partial J}{\partial\boldsymbol{\phi}_d}$ is done following \cite{Mugnier-a-01}:
first, we calculate the gradient of $J^{\text{f}}$ with respect to the PSF
$\boldsymbol{h}_c$:
\begin{equation}
\frac{\partial J^{\text{foc}}}{\partial\boldsymbol{h}_c^{\text{foc}}}=
\frac{1}{{\boldsymbol{\sigma}_n^{\text{foc}}}^2}[\alpha\boldsymbol{h}_{\text{det}}(\alpha.\boldsymbol{h}_{\text{det}}\star\boldsymbol{h}_c^{\text{foc}}-\boldsymbol{i}_c^\text{foc})].
\end{equation}

Then, the calculation consists in derivating the gradient of the PSF
$\boldsymbol{h}_c$ with respect to phases $\boldsymbol{\phi}_u[k]$ and
$\boldsymbol{\phi}_d[l]$ at pixels $k$, $l$ in pupils upstream and downstream
of the coronagraph, respectively, and applying the chain rule, as already done
in a non-coronagraphic case, e.g. in \cite{Thiebaut-a-95}. The calculation of
both gradients $\frac{\partial
  J^{\text{foc}}}{\partial\boldsymbol{\phi}_u[k]}$ and $\frac{\partial
  J^{\text{foc}}}{\partial\boldsymbol{\phi}_d[l]}$ gives

\begin{equation}
\begin{aligned}
\frac{\partial
  J^{\text{foc}}}{\partial\boldsymbol{\phi}_u[k]}&=
2\Im\left\{\boldsymbol{\psi}^{*}[k]\left[\text{FT}\left(\frac{\partial
        J^{\text{foc}}}{\partial\ \boldsymbol{h}_c^{\text{foc}}}(\boldsymbol{\Psi}-\eta_0\boldsymbol{\Psi_d})\right)\right]\right\}[k]\\
&-2\Re\left(\frac{\partial\eta_0}{\partial\boldsymbol{\phi}_u[k]}\sum_{t}\frac{\partial J^{\text{f}}}{\partial\ \boldsymbol{h}_c^{\text{foc}}}\boldsymbol{\Psi}^{*}\boldsymbol{\Psi_d}\right)\\
&+\frac{\partial|\eta_0|^2}{\partial\boldsymbol{\phi}_u[k]}\sum_{t}\frac{\partial J^{\text{f}}}{\partial\
  \boldsymbol{h}_c^{\text{foc}}}|\boldsymbol{\Psi_d}|^{2}\\
\end{aligned}
\end{equation}
\begin{equation}
\begin{aligned}
\frac{\partial J^{\text{foc}}}{\partial\boldsymbol{\phi}_d[l]}=
2\Im\bigg(&(\boldsymbol{\psi}^{*}[l]-\eta_0^{*}\boldsymbol{\psi_d}^{*}[l])\\
&\times\left\{\text{FT}\left[\frac{\partial J^{\text{f}}}{\partial\ \boldsymbol{h}_c^{\text{foc}}}(\boldsymbol{\Psi}-\eta_0\boldsymbol{\Psi_d})\right]\right\}[l]\bigg).\\
\end{aligned}
\end{equation}

With $\Im$ and $\Re$ the imaginary and real part (respectively), and
\begin{equation}
\begin{aligned}
\frac{\partial\eta_0}{\partial\boldsymbol{\phi}_u}&=j\boldsymbol{P}_u^2e^{j\boldsymbol{\phi}_u}\\
\boldsymbol{\psi}(\boldsymbol{\phi}_u,\boldsymbol{\phi}_d)&=\boldsymbol{P}_de^{j(\boldsymbol{\phi}_u+\boldsymbol{\phi}_d)} \qquad \boldsymbol{\Psi}(\boldsymbol{\phi}_u,\boldsymbol{\phi}_d)=\text{FT}^{-1}(\boldsymbol{\psi})\\
\boldsymbol{\psi_d}(\boldsymbol{\phi}_d)&=\boldsymbol{P}_de^{j\boldsymbol{\phi}_d} \qquad \boldsymbol{\Psi}_d(\boldsymbol{\phi}_d)=\text{FT}^{-1}(\boldsymbol{\psi_d}).
\end{aligned}
\end{equation}

Since the phases are expanded on a Zernike basis, we need the gradients of
$J^{\text{foc}}$ with respect to the Zernike coefficients $a_{x_i}$ of phase
$\boldsymbol{\phi}_x$. These gradients are given by the expression
\citep{Mugnier-a-01}:
\begin{equation}
\frac{\partial J^{\text{foc}}}{\partial a_{x_i}} =
\sum_{k}\frac{\partial J^{\text{foc}}}{\partial \boldsymbol{\phi}_x[k]}Z_i[k].
\end{equation}

Flux $\alpha$ and constant background $\beta$ are also analytically estimated
during the minimization, considering that
\begin{equation}
J^{\text{p}}[t]=\frac{1}{2}\sum_{t} \left |\frac{-\boldsymbol{i}_c^{\text{p}}[t] 
+ \alpha.\boldsymbol{h}_{\text{det}}[t]\star\boldsymbol{h}_c^{\text{p}}[t]+\beta}{\boldsymbol{\sigma}_n^{\text{p}}[t]} \right |^2
\end{equation}
Where p is for ``foc'' (focused) or ``div'' (diverse). For the sake of
simplicity, we shall omit the variable $t$. We have
\begin{equation}
\begin{aligned}
\frac{\partial J^{\text{p}}}{\partial \alpha}&=
\alpha\sum\frac{(\boldsymbol{h}_{\text{det}}\star\boldsymbol{h}_c^{\text{p}})^2}{{\boldsymbol{\sigma}_n^{\text{p}}}^2}
+\beta\sum\frac{\boldsymbol{h}_{\text{det}}\star\boldsymbol{h}_c^{\text{p}}}{{\boldsymbol{\sigma}_n^{\text{p}}}^2}\\
&-\sum\frac{(\boldsymbol{h}_{\text{det}}\star\boldsymbol{h}_c^{\text{p}})\boldsymbol{i}_c^{\text{p}}}{{\boldsymbol{\sigma}_n^{\text{p}}}^2}\\
\frac{\partial J^{\text{p}}}{\partial \beta}&=
\alpha\sum\frac{\boldsymbol{h}_{\text{det}}\star\boldsymbol{h}_c^{\text{p}}}{{\boldsymbol{\sigma}_n^{\text{p}}}^2}
+\beta\sum\frac{1}{{\boldsymbol{\sigma}_n^{\text{p}}}^2}
-\sum\frac{\boldsymbol{i}_c^{\text{p}}}{{\boldsymbol{\sigma}_n^{\text{p}}}^2}
\end{aligned}
\end{equation}
Which gives us, in a matricial form:
\begin{equation}
\begin{aligned}
\begin{pmatrix}
\sum\frac{(\boldsymbol{h}_{\text{det}}\star\boldsymbol{h}_c^{\text{p}})^2}{{\boldsymbol{\sigma}_n^{\text{p}}}^2} &
\sum\frac{\boldsymbol{h}_{\text{det}}\star\boldsymbol{h}_c^{\text{p}}}{{\boldsymbol{\sigma}_n^{\text{p}}}^2}\\
\sum\frac{\boldsymbol{h}_{\text{det}}\star\boldsymbol{h}_c^{\text{p}}}{{\boldsymbol{\sigma}_n^{\text{p}}}^2} &
\sum\frac{1}{{\boldsymbol{\sigma}_n^{\text{p}}}^2}\\
\end{pmatrix}
\begin{pmatrix}
\alpha\\
\beta\\
\end{pmatrix}
=
\begin{pmatrix}
\sum\frac{(\boldsymbol{h}_{\text{det}}\star\boldsymbol{h}_c^{\text{p}})\boldsymbol{i}_c^{\text{p}}}{{\boldsymbol{\sigma}_n^{\text{p}}}^2}\\
\sum\frac{\boldsymbol{i}_c^{\text{p}}}{{\boldsymbol{\sigma}_n^{\text{p}}}^2}\\
\end{pmatrix}
.
\end{aligned}
\end{equation}

A simple matrix inversion gives us the analytical estimation of the flux $\alpha$
and the background $\beta$ for each iteration.

\section{Tip-tilt estimation downstream of the coronagraph}
\label{tt_down}

\begin{figure}
\centering
\begin{tabular}{cc}
\includegraphics[width = 0.35\linewidth]{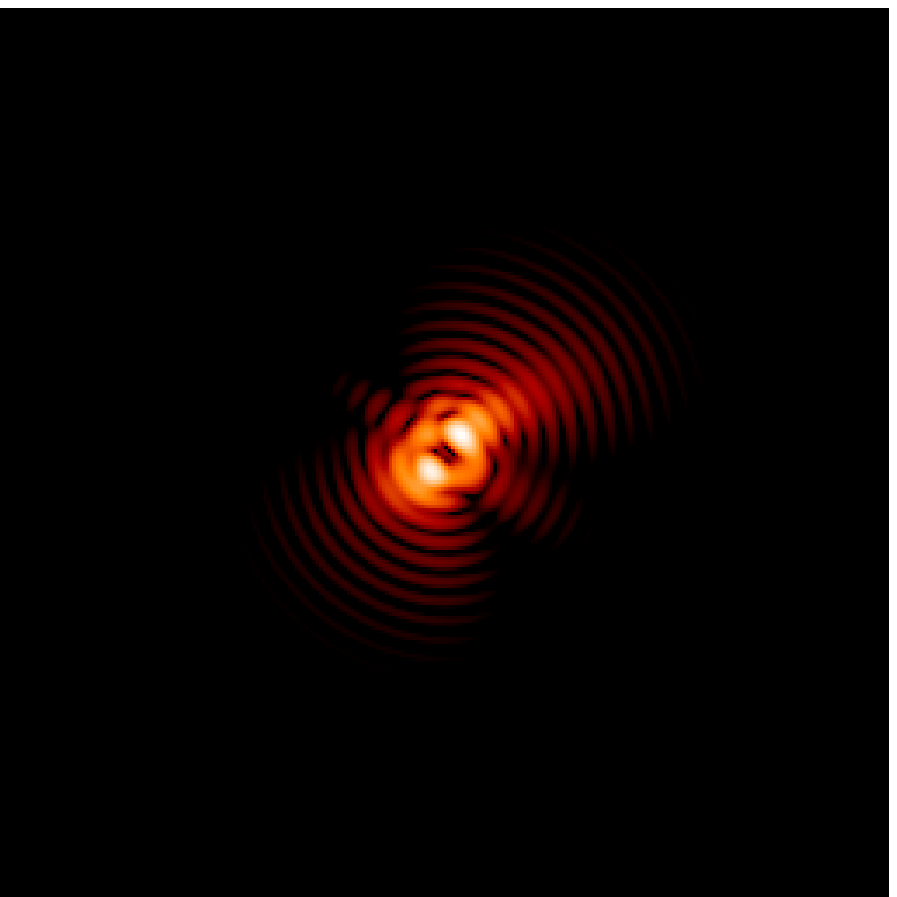}&
\includegraphics[width = 0.35\linewidth]{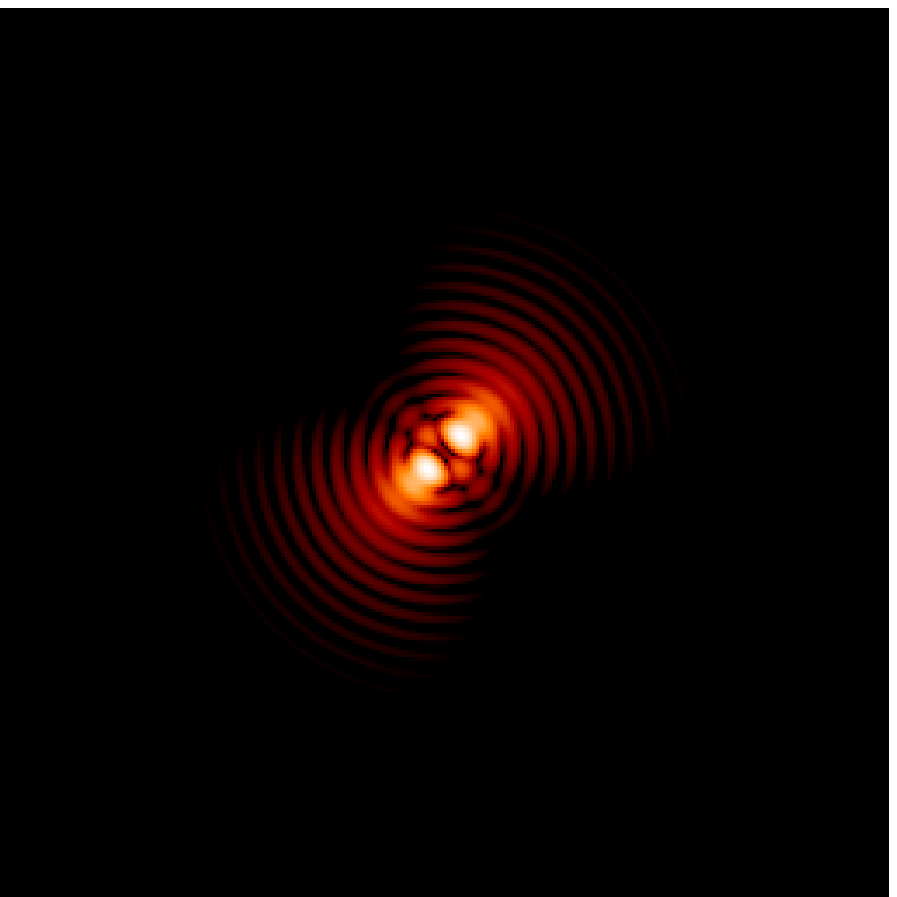}\\
\end{tabular}
\caption{Coronagraphic diversity images computed for an aberration
  $\boldsymbol{\phi}_u+\boldsymbol{\phi}_{div}$ upstream,
  $\boldsymbol{\phi}_d$ downstream of (left) and the only
  diversity aberration $\boldsymbol{\phi}_{div}$ (right). The shape of both
  images is mainly driven by diversity aberration.}
\label{imgs_div}
\end{figure}

The tip-tilt downstream of the coronagraph (which represents the image
position on the detector) strongly limits COFFEE's performance. Indeed, we
determine that the phase estimation was accurate when $-100\ \text{nm RMS}
\leq a_i \leq 100\ \text{nm RMS}$, with $a_i$ the Zernike coefficient for tip
or tilt ($i \in \left\{2,3\right\}$). Beyond this range, COFFEE is unable to
properly estimate both phases $\boldsymbol{\phi}_u$ and $\boldsymbol{\phi}_d$.
Such a phenomenon strongly limits COFFEE's performance on a bench, since its
utilization requires a restrictive location of the
PSF on the detector.\\
To get rid of this limitation, we have developed a simple and fast method of
estimating the tip-tilt downstream of the coronagraph before COFFEE's
estimation, based on the diversity image. This image is created by adding a
known aberration $\boldsymbol{\phi}_{div}=a_4^{div}Z_4+a_5^{div}Z_5$
($a_4^{div}=a_5^{div}=80$ nm RMS) to $\boldsymbol{\phi}_u$. Since the
amplitude of this aberration is important ($\sigma_{\phi_{div}}=113 $ nm RMS),
the speckles we have in the coronagraphic diversity image mainly originate in
this diversity aberration. This is illustrated in Figure \ref{imgs_div}, where
we show two diversity images: one computed with randomly generated phases
$\boldsymbol{\phi}_u$ (WFE $30$ nm RMS), $\boldsymbol{\phi}_d$ (WFE $10$ nm
RMS), and another computed with no aberrations other than the diversity ones.

As one can see in Figure \ref{imgs_div}, we can clearly identify the
aberrations which originate in the diversity $\boldsymbol{\phi}_{div}$. The
principle of our method lies in the research of these well-known aberrations
(since we know the phase $\boldsymbol{\phi}_{div}$ we introduce) in the
diversity image $i_{c}^{\text{d}}$ by comparing it with a theoretical
diversity image $i_{c_{th}}^{\text{d}}$, calculated with no other aberrations
than the diversity ones:
\begin{equation}
\boldsymbol{i}_{c_{th}}^{\text{d}}=\boldsymbol{h}_{\text{det}}\star \boldsymbol{h}_c(\boldsymbol{\phi}_{div},\boldsymbol{\phi}_d=0).
\end{equation}

The comparison of $i_{c_{th}}^{\text{d}}$ with $i_{c}^{\text{d}}$ is performed
using the method developed by \cite{Gratadour-a-05}, which
consists in minimizing the following criterion $J_{\text{TT}}$
\begin{equation}
J_{\text{TT}}(x,y)=\left \|\frac{\boldsymbol{i}_{c}^{\text{div}}(x_o,y_o) -
    \boldsymbol{i}_{c_{th}}^{\text{div}}(x_o,y_o) 
\star \boldsymbol{\delta}(x_o-x,y_o-y)}{\boldsymbol{\sigma}_n^{\text{div}}}\right\|^2\text{,}
\end{equation}
Where $\boldsymbol{\delta}$ is the dirac function. Minimization of
$J_{\text{TT}}$ gives us the shift $[x_M,y_M]$between both images. It is then
possible to calculate the corresponding tip ($a_2$) and tilt ($a_3$)
downstream of the coronagraph knowing the image sampling $s$:
\begin{equation}
a_2 = \frac{\pi}{2s}x_M \qquad a_3 = \frac{\pi}{2s}y_M.
\end{equation}
Finally, these estimated tip-tilt values are given to COFFEE as an input of
the minimization, and are used as initial values to begin phase
reconstruction. This method performs, on our experimental images, a fast
preliminary estimation ($\sim 1$ second for a $256\times 256$ image) of the
tip-tilt downstream of the coronagraph with an accuracy of $1.5$ nm RMS, which
is far enough, compared to the level of accuracy ($\pm 100$ nm RMS) required
by COFFEE.

\providecommand{\inpreparationname}{en pr\'eparation}
  \providecommand{\submittedname}{soumis}
  \providecommand{\acceptedname}{accept\'e pour publication}
  \providecommand{\tobepublishedname}{\`a para\^{\i}tre}
  \providecommand{\contractname}{Contrat}
  \providecommand{\conferencedatename}{Date conf\'erence~: }
  \providecommand{\patent}[2]{Brevet #1 #2}
  \providecommand{\firstabbrevname}{1\textsuperscript{\`ere} }
  \providecommand{\secondabbrevname}{2\textsuperscript{\`eme} }
  \providecommand{\thirdabbrevname}{3\textsuperscript{\`eme} }
  \providecommand{\fourthabbrevname}{4\textsuperscript{\`eme} }
  \providecommand{\fifththabbrevname}{5\textsuperscript{\`eme} }
  \providecommand{\sixththabbrevname}{6\textsuperscript{\`eme} }

\end{document}